# On the Origin of the Pluto System

**Robin M. Canup**
*Southwest Research Institute*

**Kaitlin M. Kratter**
*University of Arizona*

**Marc Neveu**
*NASA Goddard Space Flight Center/University of Maryland*

The goal of this chapter is to review hypotheses for the origin of the Pluto system in light of observational constraints that have been considerably refined over the 85-year interval between the discovery of Pluto and its exploration by spacecraft. We focus on the giant impact hypothesis currently understood as the likeliest origin for the Pluto-Charon binary, and devote particular attention to new models of planet formation and migration in the outer solar system. We discuss the origins conundrum posed by the system's four small moons. We also elaborate on implications of these scenarios for the dynamical environment of the early transneptunian disk, the likelihood of finding a Pluto collisional family, and the origin of other binary systems in the Kuiper belt. Finally, we highlight outstanding open issues regarding the origin of the Pluto system and suggest areas of future progress.

## 1. INTRODUCTION

For six decades following its discovery, Pluto was the only known Sun-orbiting world in the dynamical vicinity of Neptune. An early origin concept postulated that Neptune originally had two large moons — Pluto and Neptune's current moon, Triton — and that a dynamical event had both reversed the sense of Triton's orbit relative to Neptune's rotation and ejected Pluto onto its current heliocentric orbit (*Lyttleton*, 1936). This scenario remained in contention following the discovery of Charon, as it was then established that Pluto's mass was similar to that of a large giant planet moon (*Christy and Harrington*, 1978). However, *McKinnon* (1984) demonstrated that mutual Triton-Pluto interactions around Neptune were incapable of placing Pluto and Triton into their respective orbits, and instead proposed the now accepted origin scenario for Triton: that it was a separately formed Pluto-like object that was captured into a retrograde orbit around Neptune (e.g., *Agnor and Hamilton*, 2006; *Nogueira et al.*, 2011). This still satisfied the inferred genetic link between Pluto and Triton suggested by their similar absolute visual magnitudes and methane-ice dominated infrared spectra (implying similar bodies; see the chapter in this volume by Cruikshank et al.), and by their peculiar orbits dynamically linked to Neptune. However, it left open the question of the heliocentric origin of the Pluto-Charon binary. *McKinnon* (1989) subsequently made a compelling case that the angular momentum of the Pluto-Charon binary was most naturally explained by a giant impact between comparably sized bodies, with Charon forming from ejecta produced by this collision. Thus, it was proposed that Pluto-Charon originated from a similar type of collisional event to that which was becoming increasingly favored for the origin of Earth's Moon (e.g., *Benz et al.*, 1986, 1987, 1989).

Further discussion of the history of thought on the origin of Pluto-Charon can be found in the chapter addressing this topic in the 1997 *Pluto and Charon* volume (*Stern et al.*, 1997). Basic elements of our understanding of Pluto system origins described in that review persist today. For example, Pluto is still thought to have formed within a transneptunian disk of objects, and a giant impact remains the most widely accepted scenario for the formation of Pluto-Charon. However, in most respects our understanding has evolved considerably since 1997 due to both theoretical advances and new discoveries. Notable developments include:

- ***Strong theoretical and observational support for a formation closer to the Sun than today***. Early dynamical models suggested that Pluto's eccentric and inclined orbit in the 3:2 mean-motion resonance with Neptune could have originated either from interactions among bodies accreting near Pluto's current semimajor axis (*Levison and Stern*, 1995) or from formation of Pluto at a more interior distance, with Pluto then driven outward into its current orbit via resonant transport as Neptune's orbit expanded (*Malhotra*, 1993, 1995; *Malhotra and Williams*, 1997). Since the 1990s, there has been an explosion in the number of known transneptunian objects, and accounting for the orbital properties of the many bodies in resonance with Neptune (such as the "Plutino" bodies that, like Pluto, complete two heliocentric orbits in the same time as Neptune completes three) appears to require outer planet migration and resonant capture/transport (e.g., *Hahn and Malhotra*, 2005; *Levison and Morbidelli*, 2003; *Murray-Clay and Chiang*, 2005, 2006). Furthermore, there is increasing awareness that gravitational interactions with an early massive planetesimal disk would have caused the orbits of Saturn, Uranus, and Neptune to migrate outward early in the solar system's history (e.g., *Fernandez and Ip*, 1984; *Hahn and Malhotra*, 1999). Such migration is postulated to have led to dynamical instability and a sudden and major rearrangement of the giant planets, as described in the widely explored "Nice" model and its derivatives (e.g., *Tsiganis et al.*, 2005; *Levison et al.*, 2008; *Batygin and Brown*, 2010; *Nesvorný and Morbidelli*, 2012). Thus, developments in the past 20 years strongly favor an early expansion of Pluto's orbit to the current heliocentric distance.

- ***Much faster and earlier assembly***. It is now thought that large outer solar system planetesimals may have accreted much more rapidly than appreciated two decades ago, via gravitational collapse of clouds of millimeter- to decimeter-sized particles (dubbed "pebbles") into 100-km-class objects while solar nebula gas was still present in the first few to 10 m.y. of solar system history (e.g., *Youdin and Goodman*, 2005; *Ormel and Klahr*, 2010; *Lambrechts and Johansen*, 2012; *Johansen et al.*, 2015; *Nesvorný et al.*, 2019). This contrasts with hierarchical coagulation by purely two-body collisions that would require at least tens of millions to perhaps hundreds of millions of years to form large bodies at distances >15 AU (e.g., *Stern and Colwell*, 1997; *Kenyon et al.*, 2008; *Kenyon and Bromley*, 2012). This has modified thinking on when the progenitors of the current Pluto system could have formed.

- ***Quantitative understanding of the binary-forming impact scenario***. While the idea that the Pluto-Charon binary was the product of an impact was proposed in the late 1980s, the event was not modeled until more than a decade later. Three-dimensional hydrodynamical simulations have now identified a quite limited range of impacts capable of explaining the Charon-to-Pluto mass ratio, and revealed a new regime of impacts that produces Charon as an intact byproduct of the collision (*Canup*, 2005, 2011; *Arakawa et al.*, 2019). Such work implies that a Pluto-Charon forming impact requires a low impact velocity comparable to the mutual escape velocity, and that collisional outcome is sensitive to the differentiation states of the colliding bodies. These results provide new clues regarding the interior structure of the progenitors of Pluto and Charon and the circumstances and timing of Pluto system formation.

- ***Discovery of the four tiny moons***. In the years leading to the New Horizons flyby, repeated observations of the Pluto system led to the discovery of four moons tens of kilometers in size orbiting the Pluto-Charon binary (*Weaver et al.*, 2006; *Showalter et al.*, 2011, 2012). The outer moons have nearly circular and co-planar orbits, and they orbit Pluto in the same sense and plane as Charon, which seems to strongly suggest that they share a common origin with the binary (e.g., *Stern et al.*, 2006). Their surfaces are unusually bright and ice-rich compared with other similarly sized Kuiper belt objects (KBOs). If, as seems probable, their interiors are similarly dominated by water ice, this is consistent with the small moons having originated from debris dispersed during a Charon-forming impact (e.g., *Canup*, 2011). However, accounting for other properties of the tiny moons — notably their radially distant and near-resonant orbits — has proven elusive, and their origin remains an open and important question.

- ***In situ exploration of the Pluto system***. The New Horizons flyby vastly increased our knowledge of the physical properties of the Pluto system, providing new or much improved constraints relevant to how the system originated. The Pluto-Charon binary mass and angular momentum, as well as the individual densities of Pluto and Charon, are now well-constrained. The latter imply that Charon is somewhat more ice-rich than Pluto, but that both bodies contain roughly 60–70% rocky material by mass, with similar compositions compared with the diversity of compositions inferred across other large KBOs (e.g., *McKinnon et al.*, 2017). Crater counts indicate that Pluto, Charon, Nix, and Hydra are probably all ancient bodies, with oldest surface ages roughly estimated to ≥4 G.y. but still subject to uncertainties in the impactor population (*Robbins et al.*, 2017; *Singer et al.*, 2019).

## 2. CONSTRAINTS ON PLUTO SYSTEM ORIGIN

Basic properties of the Pluto system that must be explained by any origin scenario(s) are summarized in Fig. 1. Constraints on formation models derive from the system's dynamical and compositional properties, as well as from various lines of evidence for when the system likely formed. In this section, we focus on constraints that apply to any model of Pluto system origin; additional conditions implied by an origin via giant impact are discussed in section 5.

### 2.1. Pluto System's Heliocentric Orbit

The current heliocentric orbit of the Pluto system with semimajor axis ≈40 AU involves multiple resonances,

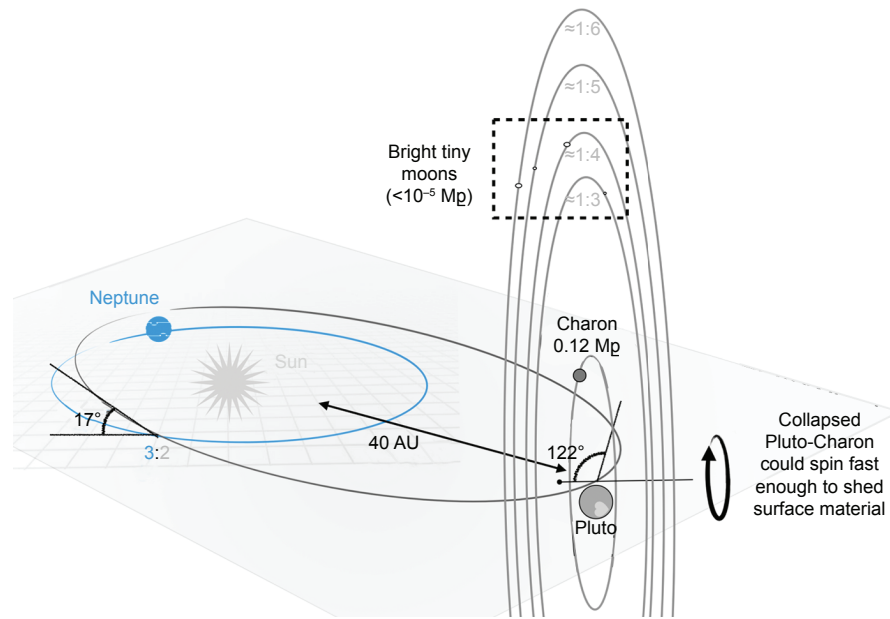

**Fig. 1.** Key constraints on the origin of the Pluto system. Pluto's heliocentric orbit is eccentric (≈0.25) and inclined (≈17°) relative to the ecliptic, and in a 3:2 mean-motion resonance with Neptune's orbit. The axes of the mutual Pluto-Charon orbit and the orbits of the tiny moons are aligned with Pluto and Charon's spin axes and form a 122° angle with the system's heliocentric orbital axis. The system's angular momentum is so high that if the combined material of Pluto and Charon was collapsed into a single object, this object could spin fast enough to shed mass. The tiny moons orbit the binary near mean-motion resonances with the mutual orbit. They are icy and much brighter than transneptunian objects of similar size. Charon's mass is a significant fraction of that of Pluto (Charon-to-Pluto mass ratio of 0.12). Charon is slightly icier than Pluto (*Bierson et al.*, 2018), but Pluto and Charon's bulk densities are still quite similar compared to the broad range of density estimates across all large Kuiper belt objects (KBOs).

most notably with the mean motion of Neptune, in which the ratio of the orbital period of Pluto to that of Neptune is 3:2. A portion of Pluto's orbit is closer to the Sun than Neptune. However, Pluto's argument of perihelion librates about 90° so that when Pluto is closest to the Sun it is also at a great height above the ecliptic and the plane of Neptune's orbit. Additional resonances involve the relative nodal longitudes and longitudes of perihelion of Pluto and Neptune (e.g., *Malhotra and Williams*, 1997). Together these resonances produce a stable configuration in which Pluto and Neptune never approach each other in three-dimensional space, even though Pluto's orbit is thought to be chaotic (e.g., *Sussman and Wisdom*, 1988).

Pluto's orbit is highly eccentric (e ≈ 0.25) and inclined relative to the ecliptic (i ≈ 17°). *Malhotra* (1993) demonstrated that if Pluto was captured into the 3:2 resonance with Neptune as Neptune's orbit migrated outward, an initially low-eccentricity Pluto orbit would have been driven to its current high-eccentricity state as Neptune's orbit expanded by about 5 AU. This implied that both Pluto and Neptune initially formed well interior to their current semimajor axes.

As first described by *Fernandez and Ip* (1984), outward migration is driven by gravitational interactions between the outer planets and a companion planetesimal disk that likely persisted after the gas-rich solar nebula had dissipated. Scatterings of planetesimals by a giant planet cause its orbit to undergo small changes. Planetesimals scattered inward by Saturn, Uranus, and Neptune tend to encounter the closer-in, more massive giant planets, whereas those scattered outward tend to encounter less-massive giant planets or (for planetesimals scattered by Neptune) no giant planet at all. Because of this asymmetry, there is a net inward flux of planetesimals scattered by Saturn, Uranus, and Neptune and a compensating outward migration of these planets so as to conserve energy and angular momentum. Jupiter, which is more massive than the other giants combined, is the most likely to eject planetesimals out of the solar system and its orbit loses the angular momentum and energy imparted to these planetesimals. This causes Jupiter's orbit to contract. As the giant planets undergo such planetesimal-driven migration, they cross mutual resonances that can lead to orbital instability and Nice-type evolutions (e.g., *Tsiganis et al.*, 2005). Dynamical models of this process developed over the past two decades have self-consistently reproduced many of the solar system's current features (e.g., *Tsiganis et al.*, 2005; *Levison et al.*, 2008; *Nesvorný and Morbidelli*, 2012). Given the successes of these models, alternative scenarios in which Pluto acquired its heliocentric orbital properties *in situ* at 40 AU (*Levison and Stern*, 1995) are no longer favored.

The orbital properties of Pluto and other KBOs constrain the character of Neptune's migration. As detailed below, they suggest that it involved a superposition of slow, continuous orbital expansion over tens of millions of years due to interaction with small objects in the planetesimal disk, and "grainy" changes (i.e., random and sudden but relatively small compared to the overall migration) in semimajor axis due to interactions with thousands of Pluto-sized planetesimals (*Nesvorný and Vokrouhlicky*, 2016). These characteristics do not preclude an additional postulated "jump" in Neptune's semimajor axis due to a close encounter with a fifth giant planet ultimately ejected from the solar system by Jupiter, a common outcome of Nice-type simulations (*Nesvorný and Morbidelli*, 2012).

Evidence for the slow, smooth component of migration arises from the need to explain Pluto's high orbital inclination. Unlike its high eccentricity, the inclination was not easily obtained in early models of resonant capture of an initially low-eccentricity Pluto by an outwardly migrating Neptune, due to the lack of a sufficiently strong mechanism to excite inclinations (e.g., *Hahn and Malhotra*, 2005; *Levison et al.*, 2008). Recent works consider that capture into the 3:2 resonance may instead have occurred through a high-eccentricity path, with bodies first scattered outward onto high-eccentricity, high-inclination orbits through interactions with Neptune, and subsequently trapped and stabilized into the 3:2 resonance by secular cycles (e.g., the Kozai resonance) and by the effects of Neptune's continued migration (*Gomes*, 2003; *Dawson and Murray-Clay*, 2012). Explaining the observed inclination distribution of KBOs in the 3:2 resonance — including Pluto as well as the Plutinos — implies that Neptune's migration was relatively slow, with an e-folding timescale $\geq 10$ m.y., so that there was sufficient time for scattering interactions and related dynamical processes to excite inclinations (*Nesvorný*, 2015; *Nesvorný and Vokrouhlicky*, 2016; but see also *Volk and Malhotra*, 2019).

With smooth migration alone, one would expect to find a larger fraction of KBOs in resonance with Neptune, due to highly efficient resonant capture mechanisms (e.g., *Hahn and Malhotra*, 2005; *Nesvorný*, 2015). In addition, smooth migration would cause resonant objects to have larger libration amplitudes than observed (*Gladman et al.*, 2008; *Nesvorný*, 2015). Migration and scattering models may be reconciled with observations if the planetesimal disk also contained a population of large, Pluto-sized objects. These large, discrete bodies introduce a graininess into Neptune's migration, akin to a random walk in its semimajor axis, which decreases both the efficiency of resonance capture and resulting libration amplitudes (*Nesvorný and Vokrouhlicky*, 2016). Such models imply that there were thousands of Pluto-class objects interior to about 30 AU as Neptune began to migrate and Pluto achieved its current heliocentric orbit. Only a small fraction ($\sim 10^{-3}$) of the disk planetesimals were then ultimately incorporated into the Kuiper belt, with most lost to accretion by the giant planets or ejection (e.g., *Nesvorný*, 2015).

### 2.2. Dynamics of the Pluto-Charon Binary

Any scenario for the formation of the Pluto system must be able to reproduce its atypically high obliquity and angular momentum (Fig. 1). The plane of the system has a 122° obliquity relative to its heliocentric orbit, so that Pluto's rotation and the orbits of all its moons are retrograde with respect to the sense of its heliocentric orbit. Pluto and Charon are on a close, tidally locked, and essentially circular orbit that is coplanar with the equators of both bodies. Their semimajor axis is only $a \approx 16.5$ Pluto radii, where Pluto's radius is $R_P \approx 1188$ km (*Nimmo et al.*, 2017), with an upper limit on the orbital eccentricity of $e \sim 10^{-5}$ (*Brozović et al.*, 2015). Together, these features indicate that the binary is in tidal equilibrium. In this dual synchronous state, the rotational day of both bodies is equal to their mutual orbital period, $P = 6.3872$ days.

Tidal circularization timescales, which increase dramatically in wider orbits, suggest that the binary has always been close together (*Farinella et al.*, 1979; *Dobrovolskis et al.*, 1997). This would imply that Pluto and Charon formed closer to one another and moved apart due to transfer of angular momentum from Pluto and Charon's initially faster spins into the initially shorter-period mutual orbit (e.g., *Renaud et al.*, 2021). The innermost distance at which Charon could have formed would be near the Roche limit, which is at $\approx 2.5\ R_P$ for Pluto and Charon's densities. During tidal expansion of the mutual orbit, tides raised on Pluto and Charon by one another could have either increased or decreased the orbital eccentricity depending on the relative strength of tidal dissipation associated with combinations of spin and orbital frequencies at any given time (*Renaud et al.*, 2021). The pace of orbital expansion, eccentricity change, and spin synchronization could have quickened or stalled in the vicinity of spin-orbit resonances. Eventually, the orbit became circular as the dual-synchronous state was reached (e.g., *Cheng et al.*, 2014a).

The Pluto-Charon secondary-to-primary mass ratio (q) is the largest of any known planet or dwarf planet satellite in the solar system, with $q = 0.122 \pm 0.0014$ (e.g., *Stern et al.*, 2018). The mean semimajor axis is $a = 19596$ km in the Plutocentric frame, which together with the orbital period implies a total system mass $M_{PC} = (1.462 \pm 0.002) \times 10^{25}$ g (*Brozović et al.*, 2015). The total system angular momentum about the center of mass, due to Pluto and Charon's spins and their mutual orbit, is

$$\begin{aligned}
L_{PC} &= \omega K_P M_P R_P^2 + \omega K_C M_C R_C^2 + \omega \frac{M_C M_P}{M_{PC}} a^2 \\
&= \frac{\sqrt{GM_{PC}^3 R_P}}{(1+q)} \left(\frac{a}{R_P}\right)^{1/2} \left[K_P\left(\frac{R_P}{a}\right)^2 + qK_C\left(\frac{R_C}{a}\right)^2 + \frac{q}{(1+q)}\right]
\end{aligned} \quad (1)$$

Here, $\omega = \sqrt{GM_{PC}/a^3}$ is the angular frequency of Pluto's spin, Charon's spin, and the binary's orbit. $K_P$ and $K_C$ are the moment of inertia constants for Pluto and Charon, defined from the moment of inertia I as $K = I/(MR^2)$ with $K = 0.4$ for a homogeneous sphere and $K < 0.4$ for a sphere with a central concentration of mass (e.g., due to differentiation into a rocky core and icy mantle). The system angular momentum is dominated by the final orbital angular momentum term in equation (1). Scaling by the quantity $L' \equiv M_{PC} R_{PC}^2 (GM_{PC}/R_{PC}^3)^{1/2} = (GM_{PC}^3 R_{PC})^{1/2}$, where $R_{PC}$ is the radius of a body with Pluto's mean density that contains the total mass of Pluto and Charon, gives

$$J_{PC} \equiv \frac{L_{PC}}{L'} = \frac{1}{(1+q)^{7/6}} \left(\frac{a}{R_P}\right)^{\frac{1}{2}} \left[ K_P \left(\frac{R_P}{a}\right)^2 + q K_C \left(\frac{R_C}{a}\right)^2 + \frac{q}{(1+q)} \right] \quad (2)$$

For a broad range of moment of inertia constants reflecting highly differentiated to undifferentiated interiors for Pluto and Charon (i.e, $0.3 \leq K_P, K_C \leq 0.4$), the scaled system angular momentum given by equation (2) falls in the range $0.386 \leq J_{PC} \leq 0.394$. A single body with moment of inertia constant K that is rotating fast enough to begin to shed mass (i.e, with centripetal acceleration close to or exceeding the gravitational acceleration at the surface) would have a scaled angular momentum J = K, with K = 0.4 being the upper limit corresponding to a uniform-density body. Thus, the Pluto-Charon system angular momentum is comparable to or greater than that in a body containing the total system mass rotating at the mass ejection limit. The high-angular-momentum budget is consistent with the presence of an even more radially extended small satellite system, as we discuss below.

### 2.3. Existence and Dynamics of the Four Small Circumbinary Moons

The system's origin(s) is further constrained by the four small moons — Styx (diameter axes 16 × 9 × 8 km), Nix (50 × 35 × 33 km), Kerberos (19 × 10 × 9 km), and Hydra (65 × 45 × 25 km) in order of increasing orbital distance (*Weaver et al.*, 2016) — discovered in the decade prior to the New Horizons flyby. The small moons, of combined mass ≈6 × 10⁻⁶ times that of the binary (*Youdin et al.*, 2012; *Brozović et al.*, 2015), orbit it on circular, coplanar, and prograde orbits that are close to 3:1, 4:1, 5:1, and 6:1 mean-motion resonances with Charon. However, the moons are not actually in these resonances; indeed, because the Pluto-Charon orbit is so nearly circular, the predicted N:1 resonance widths are vanishingly narrow (*Mardling*, 2013; *Sutherland and Kratter*, 2019). In Pluto radii, the small moons' semimajor axes lie between approximately 36 and 55 $R_P$. The moon rotations are not tidally locked, and their known spin axes have high obliquities that are not aligned with those of Pluto and Charon (*Weaver et al.*, 2016). This is likely because for their large orbital radii and small sizes, tides cannot synchronize their spins in <1 G.y., longer than the timescale of perturbation by impacts (*Quillen et al.*, 2017, 2018). More details on the properties of the small moons are provided in section 4.1 and in the chapter by Porter et al. in this volume.

The New Horizons spacecraft searched for other Kerberos/Styx-sized regular moons (~$10^{16}$ kg), but none were found down to a diameter of 1.7 km (for an albedo of 0.5) and out to semimajor axes ~8 × 10⁴ km ≈ 67 $R_P$ (*Weaver et al.*, 2016). Smaller regular moons — if they exist — could be dynamically stable out to 5 × 10⁵ km ~ 400 $R_P$, i.e, ~0.1 Pluto-Charon Hill radii (*Michaely et al.*, 2017). Bodies down to 1–2 km in size (~$10^9$ kg) will be detectable with the upcoming James Webb Space Telescope, and if they exist, could help constrain the masses of the known small moons (*Kenyon and Bromley*, 2019). New Horizons also conducted an extensive search for rings comparable to those surrounding other solar system bodies, but no such structures at Pluto were seen (*Lauer et al.*, 2018). While ring debris is unstable for most orbits between the known satellites (e.g., *Youdin et al.*, 2012; *Smullen and Kratter*, 2017; *Woo and Lee*, 2018), wider-orbit debris could be long-lived. Thus, although future observations might change this picture, the Pluto system as we know it today appears dust-free and quite compact, with outermost Hydra having nearly an order-of-magnitude smaller orbital radius than that of the largest possible bound orbit.

### 2.4. Densities and Compositions of Pluto, Charon, and the Small Moons

Origin models must account for the observed compositions of members of the Pluto system, in particular the ratio of ice to rock (silicates, metals, and carbonaceous material) in their interiors as inferred from their bulk densities and/or surface compositions. Abundances of volatile ices provide additional constraints.

The bulk densities of Pluto and Charon differ sufficiently as to imply different bulk compositions. Charon's bulk density [1700 kg m⁻³ (*Nimmo et al.*, 2017)] is lower than that of Pluto (1854 kg m⁻³), implying that Charon is likely somewhat icier; porosity alone is unlikely to fully account for this difference in density (*McKinnon et al.*, 2017; *Bierson et al.*, 2018). Because the bulk densities of Pluto and Charon are higher than those of (even compressed) ice, their interiors must include denser "rocky" material. This rock is likely mostly comprised of silicates, metals, and carbonaceous material (*McKinnon et al.*, 2008) by analogy with carbonaceous chondrites (e.g., *Howard et al.*, 2011), interplanetary dust particles, and non-ice material detected in the outer solar system such as in the Saturn system (*Hsu et al.*, 2015; *Waite et al.*, 2018; *Postberg et al.*, 2018; *Tiscareno et al.*, 2019) and comets (*Hanner and Bradley*, 2004; *Ishii et al.*, 2008). For a composition of water ice and solar-composition rock, their bulk densities imply that Pluto has roughly two-thirds rock and one-third ice by mass, while Charon has 60% rock and 40% ice (*McKinnon et al.*, 2017).

Although the bulk densities of Pluto and Charon differ, this difference is small compared with the range of densities observed across large KBOs. Density estimates for KBOs with diameters >500 km vary from ~0.8 g cm⁻³ (for 2002 UX$_{25}$) to ≥2.5 g cm⁻³ [for Eris (e.g., *McKinnon et al.*, 2017)]. When viewed in this context, Pluto and Charon's densities and bulk compositions are quite similar.

The current degree of internal differentiation of Pluto and Charon is uncertain. The New Horizons mission was not designed to perform gravity measurements needed to infer moments of inertia to constrain their internal structures. Heating by the energy of their accretion, the decay of long-lived radionuclides, and mutual tides (until their spins and mutual orbit synchronized and their mutual orbit became circular) would have likely caused the interiors of Pluto and Charon to have reached the melting temperature of ice, allowing the settling of denser rock to form a core. Thus, interior models typically assume fully differentiated current structures (e.g., *Robuchon and Nimmo*, 2011; *Hammond et al.*, 2016; *Kamata et al.*, 2019).

It is possible (especially for Charon) that rock-ice separation never occurred in the frigid outer layers, where cold rock and ice may not move even over billions of years. The configuration of a denser, undifferentiated crust atop a lower-density ice mantle produced through interior differentiation (e.g., driven by radiogenic heating) is formally gravitationally unstable, but this configuration may persist for the age of the solar system (*Rubin et al.*, 2014). Moreover, differentiation by settling of rock through melted ice may have been impeded by convective homogenization in a liquid or solid mantle. Differentiation may also be inefficient if the rock is predominantly fine-grained as observed in carbonaceous chondrite matrix and interplanetary dust particles, both thought to be good proxies for the non-ice material accreted by outer solar system bodies (*Bland and Travis*, 2017; *Neveu and Vernazza*, 2019).

The above caveats notwithstanding, geological features on both Pluto and Charon seem most consistent with significant differentiation (e.g., *Stern et al.*, 2015; *Moore et al.*, 2016). For example, their surfaces lack compressional features that would be expected for undifferentiated interiors due to the formation of denser, high-pressure ice II at depth as the bodies cooled (e.g., *McKinnon et al.*, 1997; 2017; *Hammond et al.*, 2016). Instead, extensional tectonic features are observed that are most easily explained by differentiated interiors in which global expansion occurred as early water oceans froze, forming an ice I shell (e.g., *Stern et al.*, 2015; *Moore et al.*, 2016; *Beyer et al.*, 2017). What the current interiors of Pluto and Charon tell us about the differentiation state of the progenitor bodies in an impact origin remains somewhat uncertain. This proves to be an important issue, because impact outcome depends strongly on the progenitor interior states, as discussed in section 3.3 and Fig. 3.

The masses of the tiny moons, and therefore their bulk densities, were unconstrained by New Horizons data (*Stern et al.*, 2015; *Weaver et al.*, 2016), and therefore knowledge of their compositions is based to date on observations of their surfaces. The small moons have estimated geometric albedos ranging from 0.55 to 0.8 (*Weaver et al.*, 2016), which are much higher (*Stern et al.*, 2018) than those of similarly sized KBOs, whose albedos are typically between 0.02 and 0.2 (*Johnston*, 2018, and references therein; *Stern et al.*, 2019). The small moon albedos imply ice-rich surfaces, consistent with near-infrared spectra of Nix, Hydra, and Kerberos that show absorption bands associated with crystalline water ice, as well as the presence of an ammoniated species on Nix and Hydra (*Cook et al.*, 2018).

Whether the small moons contain any significant amount (more than a few percent) of non-ice material is unknown given their unconstrained bulk density and porosity. However, it seems most probable that the interiors of the tiny moons, like their surfaces, are predominantly water ice. Based on Nix and Hydra crater counts (see section 2.5), the small moons seem to be ancient, with surfaces that have been subject to billions of years of impacts by darker exogenic material. Such impacts would be primarily erosive: Characteristic encounter velocities with the Pluto system ($\sim$1–2 km s$^{-1}$, i.e, the dispersion of Keplerian velocities in the Kuiper belt) are much greater than the escape velocity from Pluto at the small moon distances ($\sim$0.15–0.2 km s$^{-1}$), so that impacting dark material would likely be lost to escaping orbits (*Weaver et al.*, 2019). Indeed, if ejecta from impacts onto the small moons had been typically retained in the Pluto system, more uniform albedos for all the satellites would result (*Stern*, 2009), contrary to the New Horizons observations. However, even erosive impacts that did not directly deliver dark material to the moons would nonetheless darken their surfaces over time if material at depth within the moons exposed by the impacts was dark and rocky. Thus, the simplest way to explain how the tiny moons have maintained bright surfaces for billions of years is to presume that their interiors are ice-rich like their surfaces (e.g., *McKinnon et al.*, 2017; *Weaver et al.*, 2019).

New Horizons data have also shed light on the distribution of volatiles on Pluto, Charon, and the small moons. Pluto displays surface $CH_4$, $N_2$, and CO ices (*Grundy et al.*, 2016), with detections of $H_2O$ "bedrock" and ammonia ices at select locations thought to have been sourced from Pluto's interior (*Dalle Ore et al.*, 2019). Its surface is red owing to organic carbonaceous material that may be partly due to atmospheric and subsequent surface chemistry starting with the photolysis of atmospheric $CH_4$ (*Grundy et al.*, 2018) and partly due to endogenic processes (*Sekine et al.*, 2017; *Cruikshank et al.*, 2019). Solid $H_2O$ blankets airless Charon, and New Horizons compositional mapping has confirmed earlier remote observations of percent-level $NH_3$-bearing species (*Grundy et al.*, 2016; *Cook et al.*, 2018), as well as an unknown dark absorber (*Buie and Grundy*, 2000; *Cruikshank et al.*, 2015).

### 2.5. Timing of Pluto System Formation

Constraints on when the Pluto system formed can be derived from several lines of reasoning based on estimated surface ages, dynamical models of planet migration and orbital instability, planet accretion models, and inferences about the limited role of heating by short-lived radiogenic elements as outer-solar-system material was assembled.

Crater counts made by the New Horizons spacecraft suggest that Pluto, Charon, Nix, and Hydra are all ancient objects with maximum surface ages $\geq$4 G.y. (e.g., *Moore et al.*, 2016; *Robbins et al.*, 2017; *Singer et al.*, 2019). Thus, these individual objects — and probably the entire Pluto

system — appear to have developed surfaces capable of retaining an impact record quite early in solar system history.

As described in section 2.1, the influence of Neptune over the system's orbit suggests that the bodies that eventually formed Pluto and Charon accreted closer to the Sun (at 15–30 AU) than today (e.g., *Levison et al.*, 2008, 2009), with their orbital expansion and excitation caused by a rearrangement of the giant planets including the outward migration of Neptune. The timing of Neptune's migration and related giant-planet orbital instability is not firmly established by dynamical models alone, because it depends on the assumed properties of the planetesimal disk at the time the solar nebula dissipated and planetesimal-driven migration began to dominate (e.g., *Gomes et al.*, 2005). Late instability models were initially advocated as a means of triggering a late heavy bombardment that produced many large basins on the Moon some ~700 m.y. after the solar system's formation (e.g., *Gomes et al.*, 2005). However, the lack of evidence so far for cometary impactors in lunar materials that date to that era appears to undermine this argument (e.g., *Joy et al.*, 2012), because comets would have dominated the impactor flux onto the Moon during the instability (e.g., *Nesvorný*, 2018).

Instead, recent works argue that planetesimal-driven migration and outer-planet instability likely occurred much earlier. A late instability after the terrestrial planets were assembled would dynamically overexcite or destabilize their orbits (e.g., *Agnor and Lin*, 2012; *Kaib and Chambers*, 2016). The simplest way to avoid this is for the instability to have occurred well before the terrestrial planets completed their accretion at ~100 m.y. An early instability may also be the most probable outcome across a broad range of plausible planetesimal disk conditions (e.g., *Deienno et al.*, 2017), and appears needed to explain the survival of the Patroclus-Menoetius binary member of Jupiter's Trojan asteroids (*Nesvorný et al.*, 2018). In combination with a slow and "grainy" migration of Neptune (as appears necessary to account for the inclination distributions and overall proportions of resonant KBOs per section 2.1), this suggests that Pluto-sized bodies had formed in the 15–30-AU region by tens of millions of years after the solar nebula dissipated (e.g., review by *Nesvorný*, 2018).

Until the last decade, it was thought that large outer-solar system bodies formed through solely two-body collisions and hierarchical coagulation. It is possible to form a few Pluto-sized objects at <30 AU in <100 m.y. through this mode of growth, and to produce an overall Kuiper belt size distribution comparable to that seen today (e.g., *Kenyon and Bromley*, 2012; *Schlichting et al.*, 2013). However, features of the Kuiper belt size distribution predicted from such models appear inconsistent with the dearth of small craters in images of Pluto and Charon obtained by New Horizons (*Stern et al.*, 2018; *Morbidelli and Nesvorný*, 2019, and references therein). More fundamentally, while hierarchical coagulation models can reproduce the current Kuiper belt, they do not appear to naturally produce the more massive and numerous population of bodies implied by planet migration models and conditions needed to account for the properties of resonant KBOs (e.g., *Nesvorný and Vokrouhlicky*, 2016; *Morbidelli and Nesvorný*, 2019).

Instead, the currently favored mechanism for the growth of large planetesimals is the streaming instability (e.g., *Youdin and Goodman*, 2005; *Johansen et al.*, 2015; *Morbidelli and Nesvorný*, 2019; *Nesvorný et al.*, 2019). Initial grains in the circumstellar disk likely aggregated into millimeter-sized pebbles due to surface sticking forces and energy loss in inelastic impacts (e.g., *Dullemond and Dominik*, 2005; *Chambers*, 2016). Streaming instability can occur due to the interactions of such pebbles with the background gaseous solar nebula. The pressure-supported gas orbits at a somewhat slower azimuthal velocity than would a particle at the same orbital radius on a purely Keplerian orbit. As such, particles orbiting in the gas experience a "headwind" as they encounter the slower orbiting gas, which causes the particle orbits to lose energy and drift radially inward. If as particles drift inward a local concentration of particles forms, this concentration will accelerate the local gas somewhat, lessening the rate of the concentration's inward drift. The concentration can then continue to grow by accreting outer particles that are drifting inward more rapidly, and as the concentration grows, its effects on the gas strengthen, allowing its drift to slow further and its growth to continue. If this positive feedback results in the local spatial density of solids becoming sufficiently high, the concentration can rapidly gravitationally collapse to form $\sim 10^2$-km-class planetesimals directly from vastly smaller pebbles (*Johansen et al.*, 2015; *Simon et al.*, 2016). This mechanism appears uniquely able to reproduce observed features such as the predominance, color similarity, and distribution of obliquities of Kuiper belt binaries (*Morbidelli and Nesvorný*, 2019; *Nesvorný et al.*, 2019), in addition to other properties of the terrestrial and giant planets (e.g., *Levison et al.*, 2015).

Pairwise collisions between planetesimals can lead to the accretion of larger bodies bound by self-gravity, but this mode of growth is extremely slow in the outer solar system. If instead planetesimals grow primarily via the accretion of pebbles in the presence of the gas disk, a dramatically faster mode of growth can ensue. In so-called "pebble accretion," pebbles within a preferred size range are slowed by gas drag as they pass in close proximity to a planetesimal. The drag causes pebbles that would otherwise be scattered away to instead spiral into the planetesimal and be accreted, greatly increasing the planetesimal's effective accretional cross-section and its growth rate.

Models suggest that $10^2$-km bodies could have formed near 25 AU via streaming instability in as little as a few million years (e.g., *Ormel and Klahr*, 2010, *Johansen et al.*, 2015; *Lambrechts and Morbidelli*, 2016), with continued growth via pairwise collisions and pebble accretion producing in ≈3 m.y. a large-body size distribution consistent with that needed to explain current Kuiper belt properties [e.g., with a thousand Pluto-scale objects (*Morbidelli and Nesvorný*, 2019)], assuming continued presence of the background gas disk. When conditions favorable to streaming instability actually occurred in the outer solar system is uncertain. The instability requires an enhanced solids-to-gas ratio, which could be achieved as the gas disk started to dissipate or earlier if there were solid-concentrating mechanisms such as settling and/or radial drift (e.g., *Johansen et al.*, 2009; *Morbidelli and Nesvorný*, 2019, and references therein). Dating of meteorite phases that form in the presence of gas indicates that the gas disk beyond Jupiter lasted at least 3–5 m.y. after the "time zero" of solar system formation, defined as the time of condensation of calcium-aluminum-rich inclusions (CAIs) (*Russell et al.*, 2006; *Desch et al.*, 2018), arguing for large-body growth via streaming instability and pebble accretion by this time.

In contrast, planetesimal accretion times no earlier than 5 m.y. seem implied by the dearth of aqueous and volatile processing on outer solar system small bodies and fragments thereof (i.e., interplanetary dust particles), which for earlier formation times would have resulted in intense heating from the decay of short-lived radionuclides such as $^{26}$Al, whose half-life is 0.7 m.y. (*Davidsson et al.*, 2016; *Neveu and Vernazza*, 2019). Formation after the decay of short-lived radionuclides could be a reason why at least some of Pluto's CO, which is unstable to aqueous processing (*Shock and McKinnon*, 1993), has survived to the present day. However, this argument is nuanced by the existence of aqueously altered silicate rock and of metal differentiated from silicates in meteorites linked to parent outer solar system small bodies (*Hiroi et al.*, 2001; *Kruijer et al.*, 2017), suggesting that at least some outer solar system planetesimals formed earlier than 5 m.y. In the more interior giant planet region, large planetesimals clearly formed prior to this time, as is needed to account for Jupiter and Saturn's cores and their large-scale gas accretion prior to nebular dispersal.

Taken as a group, the above arguments suggest that formation of large, $10^2$–$10^3$-km-class bodies via streaming instability + pebble accretion occurred in the 15–30-AU region within a few to 5 m.y. after CAIs. Pairwise collisions between such bodies, including a Pluto-Charon-forming impact, would have continued after nebular dispersal over the subsequent millions to tens of millions of years (see section 5.1). The considerations discussed in this section leave open the timing of the Pluto system's formation relative to Pluto's capture into the 3:2 resonance with Neptune. In the simplest case, Pluto and its moons could have formed prior to resonance capture, with the system surviving the subsequent heliocentric migration and resonant trapping (*Pires et al.*, 2015; *Nesvorný and Vokrouhlický*, 2019; *Nesvorný et al.*, 2019). Given arguments in favor of an early outer-planet instability, this suggests a Pluto-system-forming event within the first few tens of millions of years after CAIs. Conversely, if Pluto (or proto-Pluto) was trapped in the 3:2 resonance first, a hypothesized Charon-forming impact (section 3) could have dislodged the pair from resonance (*Levison and Stern*, 1995; *Hahn and Ward*, 1995; *Stern et al.*, 1997), but it could have been recaptured (e.g., *Dobrovolskis et al.*, 1997). This might allow for later Pluto-system formation times $\sim 10^8$ yr after CAIs. We return to this issue in section 5.1.

## 3. ORIGIN OF PLUTO-CHARON BY A GIANT IMPACT

Even prior to the development of quantitative impact models, the Pluto-Charon binary's low mass ratio, high angular momentum, and close separation made an impact with Pluto by a similarly sized impactor the favored origin scenario (*McKinnon*, 1984, 1989; *Stern et al.*, 1997). In the past two decades, impact simulations (*Canup*, 2005, 2011; *Arakawa et al.*, 2019) have demonstrated not only the ability of this scenario to account for such characteristics, but also the potential of a giant-impact origin to provide insight into the physical state of the progenitors and the timing of binary system formation. However, alternative origin models have also been proposed (Fig. 2).

### 3.1. Non-Giant-Impact Formation Scenarios

Fission of a fast-spinning Pluto could explain a resulting high-angular-momentum binary (*Mignard*, 1981; *Lin*, 1981). However, the amount of spinup needed to launch material into a ring from which Charon accretes (*Tancredi and Fernandez*, 1991) seems most easily achievable with a giant impact (*Stern et al.*, 1997). Indeed, some current giant-impact-origin models (e.g., Fig. 5) are akin to impact-induced fission.

Intact capture of Charon into Pluto orbit could be enabled by dynamical friction from surrounding small bodies (*Goldreich et al.*, 2002) or pebbles. This scenario tends to preferentially produce retrograde binaries (like Pluto-Charon) that are initially on widely separated orbits (*Schlichting and Sari*, 2008). Continued dynamical friction and/or scattering interactions cause the binary orbital radius to shrink, perhaps providing an alternative to the standard tidal expansion model as a means of accounting for the current Pluto-Charon dual-synchronous state. For example, dynamical friction could cause Charon's orbit to decay inward until it was balanced by the torque due to tides raised by Pluto and Charon on one another, with Charon's orbit eventually tidally expanding to the synchronous state as the small-particle disk dissipated and dynamical friction ended. However, this capture mechanism was primarily proposed to explain KBO binaries with $\sim 10^2$-km-radius objects (*Goldreich et al.*, 2002). Capture of much larger Charon appears to require an improbably dense disk (*Stern et al.*, 2018). Furthermore, it would seem difficult to explain the small moons orbiting beyond Charon in this scenario. They would presumably have to form after Charon's capture and orbital contraction through a separate, later event(s), which would not naturally explain their coplanarity and similar sense of rotation with the binary.

Co-accretion of the Pluto-Charon binary was dismissed by *Stern et al.* (1997) as not being able to account for the Pluto system's high obliquity and high angular momentum. However, the modern version of this scenario is distinct from what was termed co-accretion in older lunar origin studies (and in *Stern et al.*, 1997), owing to developments in the understanding of planetesimal formation discussed in section 2.5. Gravitational collapse induced by the streaming instability now appears to be a compelling mechanism for producing binary planetesimals up to ~$10^2$ km in size (*Nesvorný et al.*, 2019). As an azimuthal mass concentration produced by streaming instability undergoes self-gravity-driven collapse into planetesimals, the high specific angular momentum of a collapsing clump can prevent accretion into a single body and instead yield a high-angular-momentum binary (*Nesvorný et al.*, 2010). A high obliquity and mass ratio similar to those of Pluto-Charon are possible (although not the likeliest) outcomes of accretion simulations (*Nesvorný et al.*, 2019). However, direct formation of a binary as massive as Pluto-Charon by this process is unlikely given the tendency for such a large mass concentration to fragment before collapse is complete (*Youdin and Goodman*, 2005; *Li et al.*, 2019). If formed together by direct gravitational collapse, Pluto and Charon should have similar compositions, which is not quite the case although the difference in their bulk densities is small [8% (*McKinnon et al.*, 2017)]. Co-accretion of the small moons is an enticing feature of this alternative scenario, because material could potentially be initially placed near the current moon orbits, removing the need for substantial outward migration of the moons or their source material that has proven elusive in the impact scenario (section 4). However, co-accretion would produce moons with compositions similar to those of Pluto and Charon, and this appears inconsistent with their currently inferred ice-rich compositions (*Stern et al.*, 2018). Overall, it thus still appears unlikely that Pluto and Charon formed by co-accretion.

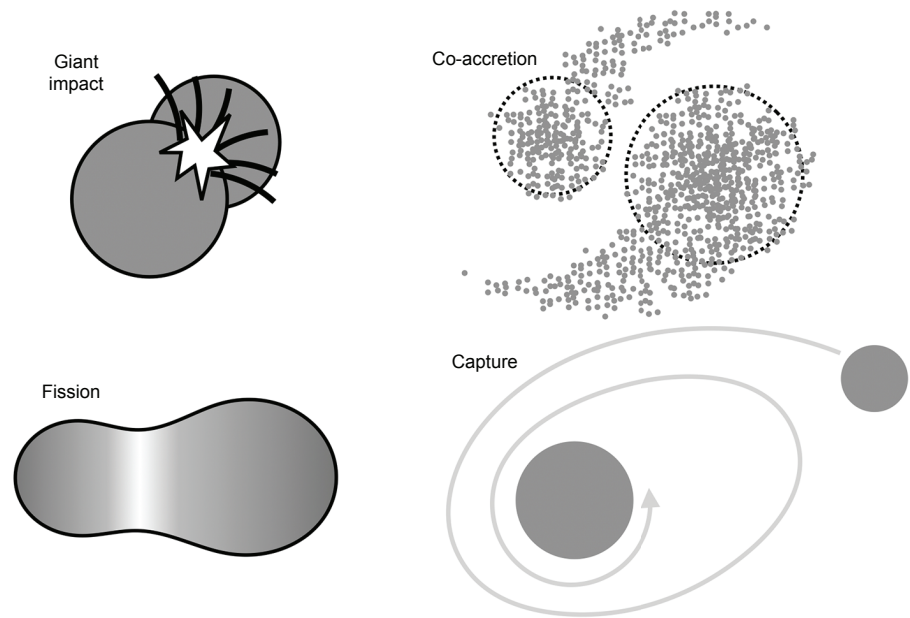

**Fig. 2.** Possible scenarios for the formation of a planetary binary. Each has been postulated as being responsible for the formation of Pluto-Charon or of other moon-bearing transneptunian objects: giant impact, currently the leading scenario for Pluto-Charon (*McKinnon*, 1984; *Canup*, 2005, 2011); fission (*Lin*, 1981; *Mignard*, 1981); co-accretion (*Nesvorný et al.*, 2010, 2019); and capture (*Goldreich et al.*, 2002).

### 3.2. Dynamical Constraints on a Pluto-Charon-Forming Impact

Consider that a collision produces the current binary masses and angular momentum. The scaled angular momentum delivered by a single giant impactor of mass $M_i = \gamma M_T$, where $M_T$ is the total colliding mass, is

$$J_i = \sqrt{2}\ f(\gamma)\ b\left(\frac{v_i}{v_{esc}}\right) \quad (3)$$

where $b = \sin \xi$ is a scaled impact parameter, $\xi$ is the impact angle (with $b = 1$ and $\xi = 90°$ corresponding to a grazing impact), $v_i$ and $v_{esc}$ are the impact and mutual escape velocities, respectively, and $f(\gamma) = \gamma(1-\gamma)\sqrt{\gamma^{1/3} + (1-\gamma)^{1/3}}$ (e.g., *Canup et al.*, 2001). With $v_i/v_{esc} \approx$ unity, which minimizes the mass and angular momentum ejected onto escaping orbits by the collision, the maximum angular momentum delivered by the grazing impact of equal-sized bodies ($\gamma = 0.5$) would be $J_{i,max} \approx 0.45$; a similar impact with $\gamma = 0.3$ would yield $J_{i,max} \approx 0.37$. Per equation (2), the Pluto-Charon binary has $0.386 \leq J_{PC} \leq 0.394$. Thus, a highly oblique impact between bodies comparable in size to Pluto itself is needed to account for the current system angular momentum.

It is of course likely that before a Charon-forming collision, proto-Pluto and the impactor would have already been rotating due to the cumulative effects of impacts during their accretion, and their spin angular momenta would have also contributed to that of the final system after a Charon-forming impact. The rate and directionality of such pre-giant impact spins is highly uncertain; the shortest possible progenitor pre-impact day would be about 2.4 h, corresponding approximately to the limit at which a Pluto-density body would begin to shed mass.

Charon's large mass compared with that of Pluto strongly constrains the type of collision that could have produced the binary. A very broad range of collisions can produce satellites with masses between $O(10^{-3})$ to few $\times 10^{-2}$ times the primary's mass (e.g., *Canup*, 2014; *Arakawa et al.* 2019). However, only a much narrower range of collisions can produce a moon with $q \geq 0.1$. In general, producing a Charon-sized moon requires a very large, oblique impact (often in combination with some pre-impact rotation) in order to maximize the total angular momentum, together with a low impact velocity to minimize the escaping mass and maximize the retained angular momentum. Higher-impact-velocity collisions that result in substantial mass loss can form large satellites, but no cases that yield $q \geq 0.1$ have been found for impact velocities in excess of 1.2 $v_{esc}$ across many hundreds of simulations that span $1 \leq (v_i/v_{esc}) \leq 1.7$, $30 \leq \xi (°) \leq 75$ and both differentiated and undifferentiated progenitor states (*Arakawa et al.*, 2019).

### 3.3. Physical State of Proto-Pluto and Giant Impactor

The above angular momentum arguments imply collisional progenitors that are both intermediate in size to Pluto and Charon (i.e., with masses between 0.3 and 0.5 $M_{PC}$). We here consider the range of likely interior states for the target and impactor at the time of a Pluto-system-forming impact, guided by both observational constraints from current Pluto and Charon (section 2), and simplified theoretical models for progenitor accretion.

*3.3.1. Constraints on progenitor interiors based on their formation histories.* As described in sections 3.4–3.5, a key parameter that strongly affects impact outcome is the distribution of ice and rock in the progenitor interiors at the time of their collision. Possible states of ice-rock differentiation range from a homogeneous, mixed ice-rock interior to a fully differentiated interior with a rocky core overlaid by an ice shell. The degree of differentiation along this spectrum depends on the progenitors' thermal state and, potentially, on the size distribution of rock grains (Fig. 3).

The onset of ice-rock differentiation in a homogenous body requires the onset of melting of ice in order for rock to settle toward the center due to its higher density. In the absence of accretional heating, the first regions to heat up to the melting temperature of ice are at the center that is most insulated from the cold surface. The melting temperature of ice depends on its composition: It is around 273 K for pure water ice across relevant pressures, but the solidus (temperature of first melt) of $NH_3$-bearing ice is depressed to 176 K if the ice contains even small, percent-level amounts of $NH_3$ (which is the case in the Pluto system; section 2.4). Once melting ensues, the ice viscosity plummets by orders of magnitude (*Arakawa and Maeno*, 1994) to allow rock settling (*Desch et al.*, 2009). The chief interior heat source in isolated bodies is the decay of short- and long-lived radionuclides. In section 2.5 we argue against a major role for short-lived radiogenic heating in the progenitor interiors, based on a likely timescale of several million years needed for progenitor accretion (e.g., *Morbidelli and Nesvorný*, 2019) and a lack of widespread

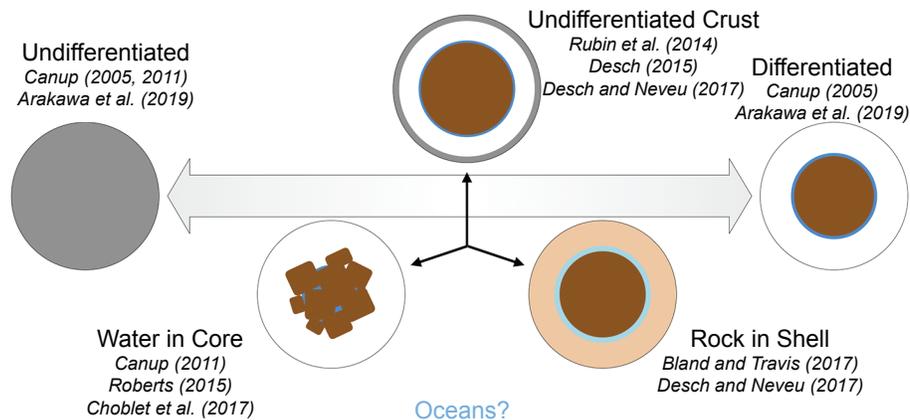

**Fig. 3.** Degeneracy in possible internal structures (drawn very roughly to scale) of the ~$10^3$-km-radius collisional progenitors that could yield Pluto and Charon with their observed densities. While a fully homogeneous structure or a fully differentiated structure (in the limit of an extremely grazing impact) can yield the Pluto-Charon binary (*Canup*, 2005; *Arakawa et al.*, 2019), a partially differentiated structure appears needed to simultaneously produce a dispersed debris disk from which the small moons could accumulate (*Canup*, 2011). However, in the spectrum between these end members, geophysical processes acting between the time of progenitor formation and that of impact can lead to mixtures of rock and water in various proportions in the core, mantle, and/or crust. The oceans may be much thicker than depicted if ≥40% of the kinetic energy of accreted planetesimals is retained (see Fig. 4e–f).

observed thermal processing of outer solar system materials (e.g., *Neveu and Vernazza*, 2019).

Even so, the buildup of heat from the decay of remaining long-lived radionuclides can by itself initiate internal ice-rock differentiation (but not rock-metal separation) within tens of millions of years of progenitor formation (Figs. 4a,b). In this mode, differentiation proceeds from inside-out. Once differentiation proceeds out to a large enough radius as to allow the growth of Rayleigh-Taylor "lava-lamp" gravitational instabilities, such instabilities can take over in overturning the denser homogeneous outer crust overlying the differentiated mantle richer in ice. Rayleigh-Taylor instabilities grow at ice-crust interfaces warmer than about 150 K, a threshold that depends modestly on other parameters (e.g., grain size) because of the tremendous dependence of viscosities on temperature (*Rubin et al.*, 2014). For formation ≥5 m.y. after CAIs and neglecting accretional heating, this 150 K threshold was likely not reached within tens of kilometers of the surface, assuming a surface temperature of ≈100 K around 20 AU and a thermal gradient of ≈1 K km$^{-1}$ typical of the insulating properties of ice at these temperatures and heat fluxes on 1000-km-class bodies. This is confirmed in numerical simulations (*Desch and Neveu*, 2017). This could lead to progenitors with an outer undifferentiated crust, thick enough to withstand bombardment by projectiles up to tens of kilometers in size until the time of the giant impact (*Desch*, 2015). This crust could comprise 10–20% of the total mass of objects 800–1000 km in radius (Fig. 4b).

Accretional heating can make differentiation occur earlier and affect the distribution of material in the progenitor's interior. Consider a simplified estimate in which some fraction of the accretional kinetic energy is symmetrically deposited at depth and retained within a growing progenitor. At each progenitor radius R, a fraction h of the impacting energy is deposited into an accreting layer, which increases the layer temperature by ΔT via the energy balance (e.g., *Stevenson et al.*, 1986, and references therein)

$$\frac{1}{2} h \dot{M} v_i^2 = C_p \Delta T \dot{M} \quad (4)$$

where $\dot{M}$ is the mass accretion rate, $v_i^2 = v_{rel}^2 + v_{esc}^2$ is the square of the impact velocity of accreting material, $v_{rel}$ is the relative velocity at large separation, and $C_P$ is the specific heat of the accreting material. The radial temperature profile due to accretional heating alone is then

$$T(R) = T_0 + \frac{4h\chi G\pi\bar{\rho}}{3C_p} R^2 \quad (5)$$

where $\chi = v_i^2/v_{esc}^2$. The (notoriously) least certain quantity in such a treatment is h. For small accreting material, h would be small too. But if, as currently thought, the progenitors grew by accreting a substantial fraction of their mass in ~10$^2$-km planetesimals formed via the streaming instability, a value h ~ tens of percent could apply. In Figs. 4c,d, we apply the *Desch and Neveu* (2017) model to a progenitor whose initial temperature profile is calculated from equation (5), with $T_0$ = 40 K (as in Figs. 4a,b), h = 0.2 (i.e., 20% of accretional kinetic energy retained as the progenitor formed), χ = 2 (i.e., a relative velocity comparable to the progenitor escape velocity), and $\bar{\rho}$ = 1.84 g cm$^{-3}$. The calculation assumes that progenitor accretion was completed at 7 m.y. post CAIs, and neglects any radiogenic heat deposited within the progenitor or its components prior to that time. Differentiation commences in the outer regions of the progenitor within less than 10 m.y., creating an outer ice shell overlying a thin water-NH$_3$ ocean. Similar outcomes are obtained with h = 0.1 if the progenitor forms earlier at 5 m.y. post-CAIs; later formation times and/or smaller h lead to evolutions similar to the no accretional heating case (Figs. 4a,b), only with somewhat earlier differentiation times. These results hold up to h = 0.35.

For h ≥ 0.4, as may be most appropriate for large planetesimal impacts, the outer regions start out melted (Figs. 4e,f). In this mode, differentiation starts during accretion and proceeds from outside-in, because temperatures initially decrease with depth (equation (5)). Although this is not captured in the model shown in this figure, rock and ice should separate in the melted outer regions, leading to a temporary interior structure with undifferentiated central regions surrounded by a denser (i.e., gravitationally unstable) rocky layer, itself enveloped by a liquid water ocean and ice shell. Differentiation by settling of rock in melted ice is complete by 30 m.y. (for h = 0.4) and potentially earlier if the gravitationally unstable rocky layer is overturned. The accretional heat can subsequently sustain an ocean more than 100 km thick until at least 60 m.y. As the ocean refreezes, there is at 60 m.y. a transition from a thin, essentially conductive ice shell above the vigorously convective ocean to a thick convective ice shell above a thin liquid layer. This decreases the overall efficiency of heat transport between the core and the surface, causing the core to warm up. Thus, this simulation suggests that for high h the structure of the Pluto-Charon progenitors at the time of the binary-forming impact would depend rather sensitively on the timing of this event relative to progenitor accretion.

*3.3.2. Degeneracies in possible progenitor internal structures.* The thermal evolution models in Fig. 3 assume that rock from zones above the melting temperature is moved instantaneously to the progenitor center. This idealized, discrete layer picture (e.g., homogeneous crust surrounding an ice mantle surrounding a rocky core, or a pure ice shell overlying a rocky core) must be nuanced by two considerations: (1) that the rocky core could be hydrated and/or retain pore ice, and (2) that the ice mantle could retain fine rock grains. Interior temperatures low enough for such structures can persist for tens of millions of years or longer, in part because much energy goes into melting ice rather than raising temperatures and in part because in regions warm enough for ice to be melted, convective heat transport by water circulation prevents temperatures from further rising significantly. These degeneracies lead to a continuum of possible structures for the progenitors (Fig. 3).

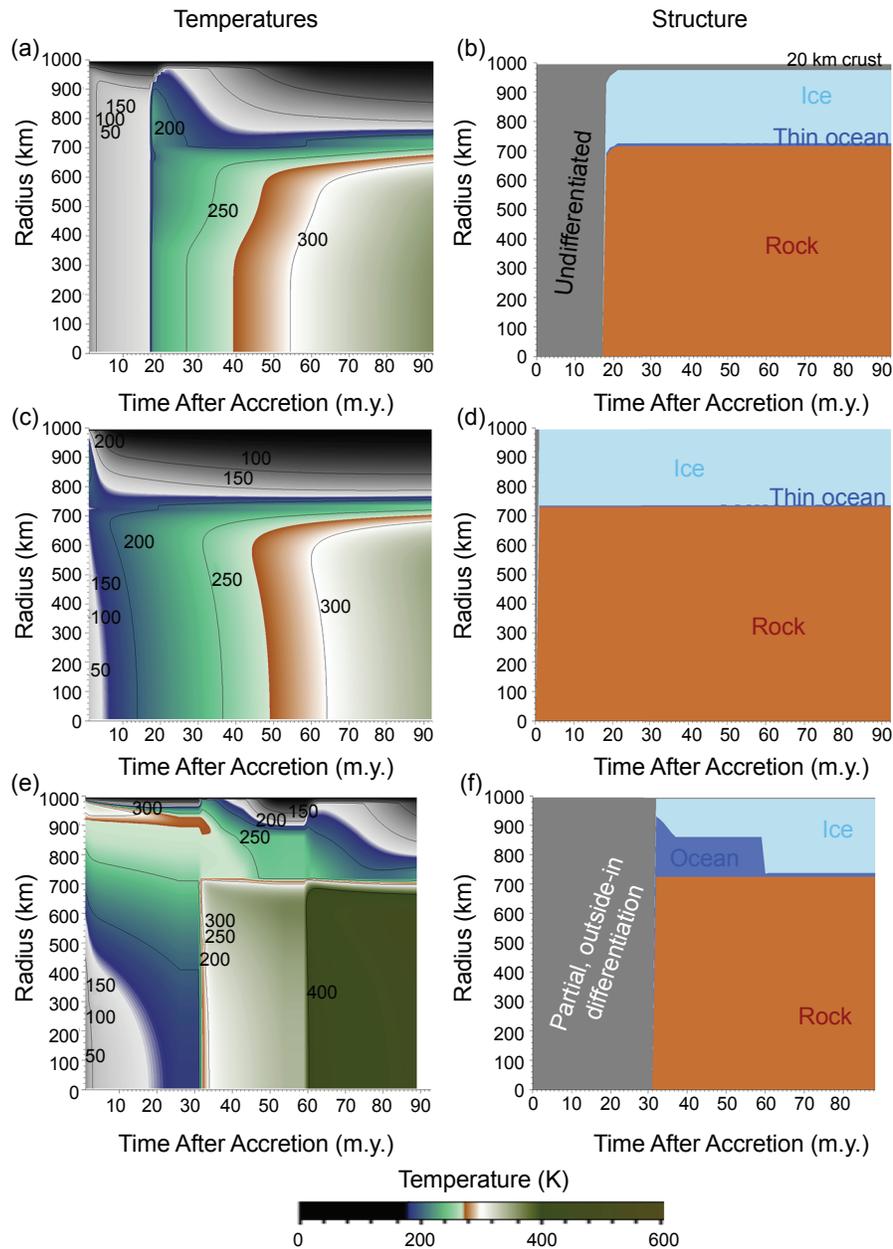

**Fig. 4.** Example thermal and structural evolution of a Pluto-Charon progenitor. **(a),(b)** Evolution of a progenitor formed 5 m.y. after Ca-Al-rich inclusions with no accretional heating. The progenitor forms cold (interior temperature = surface temperature) and homogeneous. Primarily long-lived radioactivity warms the interior [**(a)**] to trigger ice-rock differentiation, leading to a partially differentiated structure with a frigid outer ≈20 km that remains as an undifferentiated crust [**(b)**]. In **(a)** and **(c)**, contours and colors indicate temperature in Kelvin, with a transition to blue occurring at 176 K at the $H_2O$-$NH_3$ eutectic and a transition to orange indicating water ice melting at 273 K. **(c),(d)** Evolution of a progenitor formed 7 m.y. after CAIs, with an initial radial temperature profile set by accretional heating, assuming 20% of accretional kinetic energy is retained (h = 0.2; see section 3.3.1). Differentiation commences in the outer regions of the progenitor within less than 10 m.y., creating an outer ice shell overlying a water-$NH_3$ ocean. **(e),(f)** Evolution of a progenitor formed 5 m.y. after CAIs, assuming 40% of accretional energy is retained (h = 0.4). The outer layers are partially or fully melted from the outset, but because heat is more deeply buried, the warmer ice at depth conducts heat downward from these upper layers more slowly, which stalls melting of the central regions until 30 m.y. after accretion. During that time, differentiation proceeds from the outside-in within the melted layers, although the resulting advection of material and heat is not simulated here until the central layers melt. Post-differentiation, all the melted ice is in a >100-km-thick ocean that lasts another 30 m.y. until only a thin ocean persists below a convective ice shell. In the evolution models here, ammonia is assumed to be present at 1 mass% relative to $H_2O$ and depresses the melting point of ice. The rock in differentiated regions is assumed to partition 100% into the core, leading to distinct compositional layers. However, this picture may be oversimplified: The rocky core could retain pore water, and fine-grained rock may remain suspended in the ocean and/or ice mantle rather than settle down to the core. Results based on the model of *Desch and Neveu* (2017).

The presence of substantial water in a rocky core (both mixed with and bound in hydrated silicates) has been invoked to explain the surprisingly low level of ice-rock differentiation of even warm (ocean-bearing) outer-solar-system moons such as Enceladus (*Roberts*, 2015; *Choblet et al.*, 2017; *Neveu and Rhoden*, 2019), Titan (*Castillo-Rogez and Lunine*, 2010; *Iess et al.*, 2010), or Callisto (*Anderson et al.*, 1998, 2001), as well as the enhanced tidal dissipation that results for Enceladus to explain its sustained high geologic activity. The assumption that the Pluto-Charon progenitors had ice shells overlying rocky cores that retained water was made by *Canup* (2011) in models that produced both an intact Charon and a low-mass disk from which the small moons could potentially accrete (section 3.5).

That substantial rock could be present in an icy mantle has been suggested in the context of carbonaceous chondrite parent bodies (*Bland et al.*, 2013; *Bland and Travis*, 2017) and related protoplanets such as Ceres (*Neveu and Desch*, 2015; *Travis et al.*, 2018). Much of the rock in carbonaceous chondrites and interplanetary dust particles is in micrometer-sized grains that may take time to settle even if internal temperatures promote ice-rock separation upon partial or full melting of ice. Competition between settling and convective resuspension could prevent the fine grains from being segregated into the rocky core. The consequences of this degeneracy on possible progenitor structures was discussed in further detail by *Desch and Neveu* (2017).

### 3.3.3. *Summary of possible structures and implications for the giant impact model.*

The masses, ice-rock contents, and likely formation time of Pluto and Charon are consistent with a continuum of differentiated structures for the progenitors. Endmember structures include an icy mantle overlying an undifferentiated rock-ice core (*Canup*, 2011), an undifferentiated crust surrounding a differentiated interior with an ice mantle and rock core (*Desch*, 2015), and a mantle of fine-grained rock and ice above a rock core (*Desch and Neveu*, 2017), as shown in Fig. 3. In all cases, the progenitors are large enough (radius ≈ $10^3$ km) that heat from long-lived radioactive decay alone could have enabled and maintained at least a thin global $H_2O-NH_3$ ocean just above the core and/or pore liquid in the core at the time of impact. Further constraints on the progenitor interiors could arise from the small moons if they formed concurrently with the binary (which may require partially differentiated progenitors with an outer ice shell; see section 3.5), or from forward modeling of progenitor evolution up to the time of impact if the timing of the latter can be further constrained by other arguments (e.g., planet migration models).

### 3.4. Giant Impacts that Produce Massive Disks from which Charon Later Accretes

We turn now to a discussion of the types of impacts that could account for the very large Charon-to-Pluto mass ratio. To date, these events have been modeled using smoothed particle hydrodynamics (SPH) simulations that include explicit self-gravity, but that treat the colliding bodies as strengthless (*Canup*, 2005, 2011; *Sekine et al.*, 2017; *Arakawa et al.*, 2019). This may not be a good approximation for this scale of objects. However, results to date suggest that inclusion of material strength would most strongly affect impact heating (e.g., *Davies and Stewart*, 2016; *Emsenhuber et al.*, 2018), with more limited effects on overall dynamical outcome (e.g., orbiting mass and angular momentum).

A first category of potential Charon-forming impacts involves an impact between similarly sized bodies whose interiors were differentiated prior to their collision. Figure 5 shows an example SPH simulation of an impact between highly differentiated bodies, with metallic cores, rocky mantles, and outer ice shells (*Canup*, 2005). For a low-velocity collision with a total angular momentum comparable to $L_{PC}$, a common sequence occurs, with features reminiscent of those seen in simulations of rapidly rotating protostars (e.g., *Tohline*, 2002). After an initial oblique collision, the colliding bodies separate before undergoing a secondary, merging impact. During the secondary impact, the merged body forms a bar-type instability, and the higher density components migrate to the center, while from each end of the "bar," portions of the lowest-density (ice) component emanate radially outward. As the system rapidly rotates, these outer portions differentially rotate, forming trailing spiral structures whose self-gravity transports angular momentum from inner to outer regions. Ultimately the spiral structures break up to form a massive ice disk orbiting a central object, with all or nearly all of the impactor's rock and metal accreted to the primary's central core. This general type of collision was subsequently named a "graze and merge" impact (*Leinhardt et al.*, 2010).

Figure 6 shows the predicted satellite-to-primary mass ratios (q) that would result from such collisions. Here, SPH results, together with conservation of mass and angular momentum have been used to estimate the mass of the satellite that would accrete beyond the Roche radius in the optimistic limit that no material escapes from the disk, so that all initial disk material is either accreted into a single satellite or onto Pluto (e.g., *Ida et al.*, 1997). For collisions that produce final bound systems with J > 0.35, the moon masses predicted from the SPH results can be approximated by an analytic relationship (dashed line in Fig. 6) that assumes that the post-impact primary is rotating with rate $\omega = (GM_p/R_p^3)^{1/2}$ (where subscript p quantities refer to the primary), together with a moon of mass $qM_p$ orbiting at semimajor axis $a$, so that

$$J_f \approx K \left(\frac{1}{1+q}\right)^{\frac{5}{3}} + \frac{q}{(1+q)^{\frac{13}{6}}} \left(\frac{a}{R_p}\right)^{\frac{1}{2}} \quad (6)$$

The angular momentum of a body rotating at the rate at which it begins to shed mass is proportional to its moment-

of-inertia constant, K. As K in the post-impact primary is reduced (i.e., for a more highly differentiated structure), less angular momentum can be contained in its rotation and more angular momentum is instead partitioned into orbiting material for low-velocity impacts. Thus, for otherwise similar impact conditions, the mass in the disk (and q of the resulting satellite) increases as K decreases. The highly differentiated, low-K bodies considered in Fig. 5 and for most of the cases in Fig. 6 were thus chosen to maximize q. If instead the post-impact primary were less centrally condensed — e.g., if it was only partially differentiated or if it lacked a metallic core — smaller satellite masses would be expected.

A graze-and-merge collision that produces a massive disk is difficult to reconcile with Pluto-Charon for two reasons (*Canup*, 2005). First, such collisions generally yield moons with q < 0.1, and only extreme cases appear capable of producing the large Charon-to-Pluto mass ratio. These involve highly differentiated bodies (where even rock and metal have separated) and fast pre-impact rotations in the target and the impactor that are optimally aligned to contribute substantially to that of the collision itself.

Second, massive disks produced by collisions of differentiated ice-rock bodies are ice-rich (Fig. 4) (*Canup*, 2005; *Arakawa et al.*, 2019), which is inconsistent with Charon's substantial rock content. If one instead considered the collision of partially differentiated bodies in which not all the rock is in a central core, it seems probable that the resulting disks would be more rock-rich, which is potentially consistent with Charon's bulk density (*Desch*, 2015; *Desch and Neveu*, 2017). As discussed above, such a partially differentiated state may be likely if neither accreting impactors nor interior radiogenic heating effectively warmed the outer layers of proto-Pluto or the giant impactor. However, at least based on impact simulations to date (e.g., Fig. 5), production of a dispersed disk massive enough to yield Charon would still appear to be improbable for such cases, which would tend to yield a higher effective K for the post-impact Pluto and an ultimately a lower q than that of Pluto-Charon.

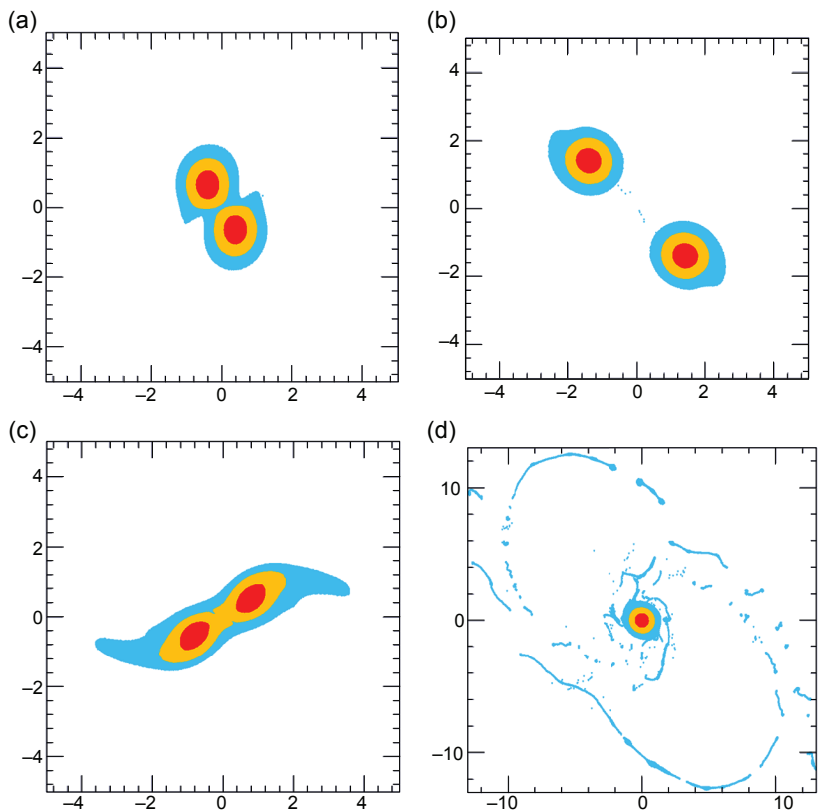

**Fig. 5.** A potential Pluto-Charon forming impact that produces a massive disk. Shown are time steps from an SPH simulation of the collision of two equal-sized, highly differentiated bodies (with iron cores in red, rock mantles in orange, and ice shells in blue) that were both rotating with a 7-h rotational day prior to the collision, with their pre-impact spin axes aligned with the collision angular momentum vector. After an initial oblique, $v_i = v_{esc}$ collision **(a)**, the bodies temporarily separate **(b)** before re-colliding and merging to form a rapidly rotating structure **(c)**. As the higher-density components migrate to the center, low-density ice expands outward and ultimately forms a massive ice disk containing 12% of the central planet's mass **(d)**. Units are distance in $10^3$ km. From *Canup* (2005).

### 3.5. Giant Impacts that Produce Charon Intact

Oblique, low-velocity impacts can also produce large intact moons, rather than a disk from which Charon later accumulates. In such cases, a large portion of the impactor remains intact after the initial impact and is torqued into a bound orbit with a periapse exterior to the Roche limit via gravitational interactions with the primary whose shape is distorted by the initial impact. This behavior was first seen in simulations by *Canup* (2005) that considered collisions between undifferentiated bodies with uniform serpentine (a hydrated silicate) composition. Figure 7 shows an example case that yields a Pluto-Charon like binary. In impacts that produce an intact Charon, Charon's bulk composition is similar to that of the bulk impactor (*Canup*, 2005; *Arakawa et al.*, 2019). Because deformation of the target strongly affects the initial torques that produce an intact moon, variations in outcome may be expected for different compositions and associated equations of state. *Canup* (2005) used the ANEOS equation of state for serpentine; broadly similar overall results were seen in *Arakawa et al.* (2019) using the Tillotson equation of state and pure-ice progenitors.

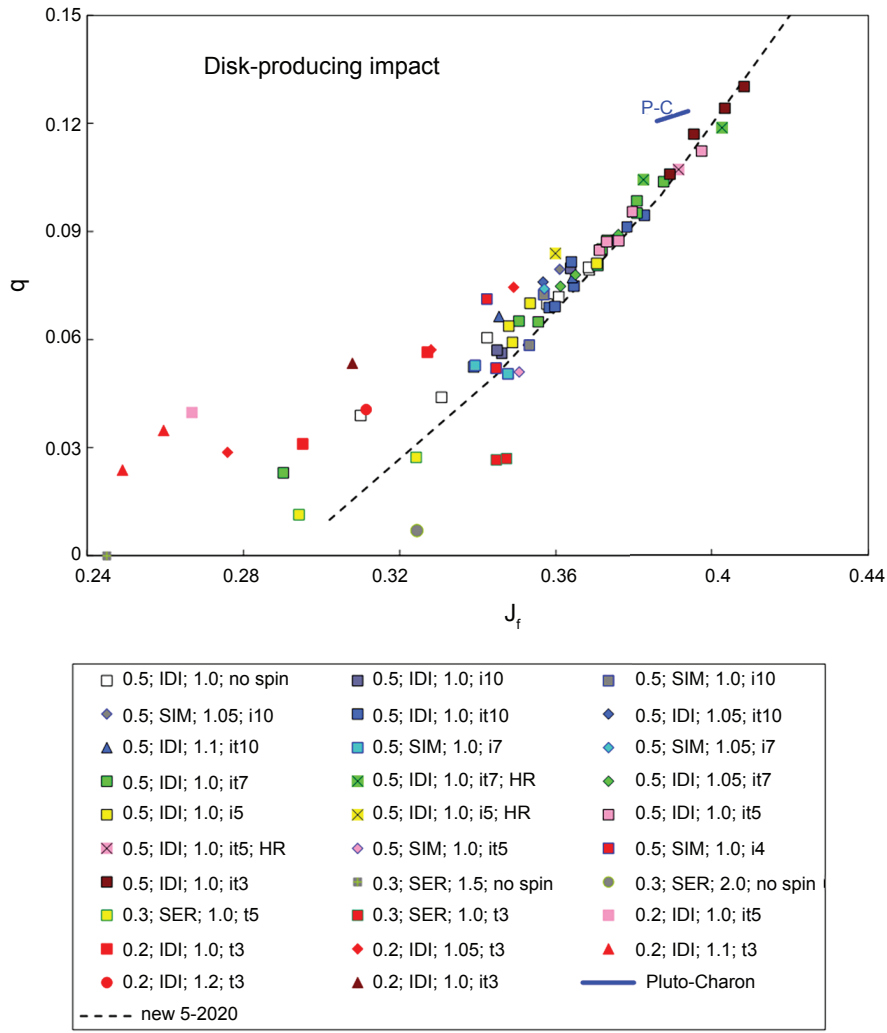

**Fig. 6.** Predicted satellite-to-planet mass ratio (q) that would result from giant impacts that produce dispersed disks as a function of the scaled angular momentum of the final bound planet + disk system ($J_f$). Legend indicates simulation parameters; the first value is the ratio of the impactor to the total mass ($\gamma$), the second indicates the composition of the colliding objects (IDI = differentiated with iron, dunite, and ice layers; SER = undifferentiated, uniform serpentine composition; SIM = undifferentiated mixture of 50% ice and 50% serpentine), the third is the ratio of the impact velocity to the escape velocity, and the fourth is the pre-impact prograde spin period in hours for the impactor ("i") and/or target ("t"), where the pre-impact spin axes were aligned with the collisional angular momentum vector. Most cases used 20,000 SPH particles; 120,000 particle simulations are indicated with "HR." The dashed line is the relationship between q and $J_f$ from equation (6) for an IDI planet of mass $M_p$ and moment of inertia constant K = 0.29, rotating with a period equal to its minimum for rotational stability and having a moon of mass $qM_p$ orbiting at a distance of about 3 primary radii. Modified figure based on *Canup* (2005), including post-New Horizons values for the Pluto-Charon system (bright blue solid line).

The formation of large intact moons is affected by the impactor's differentiation state prior to the collision. For a uniform impactor composition, corresponding to an undifferentiated hydrated silicate or perhaps a rock-ice mixture, differential motion across the impactor after the initial impact is somewhat reduced, allowing self-gravity to maintain a substantial portion of the impactor as an intact clump that becomes a Charon-analog. For a low impact velocity [i.e., $v_i/v_{esc} \leq 1.2$ (*Canup*, 2005, 2011; *Arakawa et al.*, 2019)], such collisions yield intact moons with $0.1 < q < 0.4$ across a relatively broad range of impactor masses ($\gamma \geq 0.3$), impact angles (50°–75°), and varied pre-impact spin states (Fig. 8). Producing a Pluto-Charon-like mass ratio and total angular momentum is an intermediate outcome among such cases, in contrast to the extremely limited range of impacts that appear capable of producing Charon from a dispersed disk. Thus, intact formation of Charon appears the more probable mode (*Canup*, 2005). A notable property of all such cases is that Charon's initial orbit is eccentric, with $e \sim 0.1$ up to on the order of unity (*Canup*, 2005, 2011; *Arakawa et al.*, 2019).

It is, however, possible that a differentiated impactor could also produce an intact Charon. *Arakawa et al.* (2019) performed SPH simulations of impacts between differentiated bodies with 50% of their mass in an inner rock core and 50% in an outer ice shell. For this structure, one extremely oblique (impact angle 75°) collision was identified that produced a Charon-sized intact moon (*Arakawa et al.*, 2019) (Fig. 9). However, Arakawa et al. found that Charon-sized intact moons were more probable with undifferentiated impactors, as in *Canup* (2005).

After the discovery of Pluto's four tiny moons, studies were initiated to explore whether a giant impact capable of producing Charon intact could also yield dispersed debris from which the outer moons could form. *Canup* (2011) considered collisions between partially differentiated progenitors, whose interiors were comprised of hydrated silicate (90% serpentine by mass) with an overlying water ice shell (10% by mass). This structure was intended to represent a case in which radiogenic heating was limited, so that complete separation of rock from ice had not occurred in the interior at the time of the collision, with an outer ice layer produced by impact-driven melting at the end of the progenitors' accretion (e.g., *Barr and Canup*, 2008) (Fig. 4b). The mean density of the partially differentiated progenitors in *Canup* (2011) was $\approx 2$ g cm$^{-3}$, somewhat larger than what the densities of Pluto and Charon proved to be. Similar overall results would presumably be found with lower-density progenitors having similar internal structures but thicker ice shells and/or lower-density rock-ice cores.

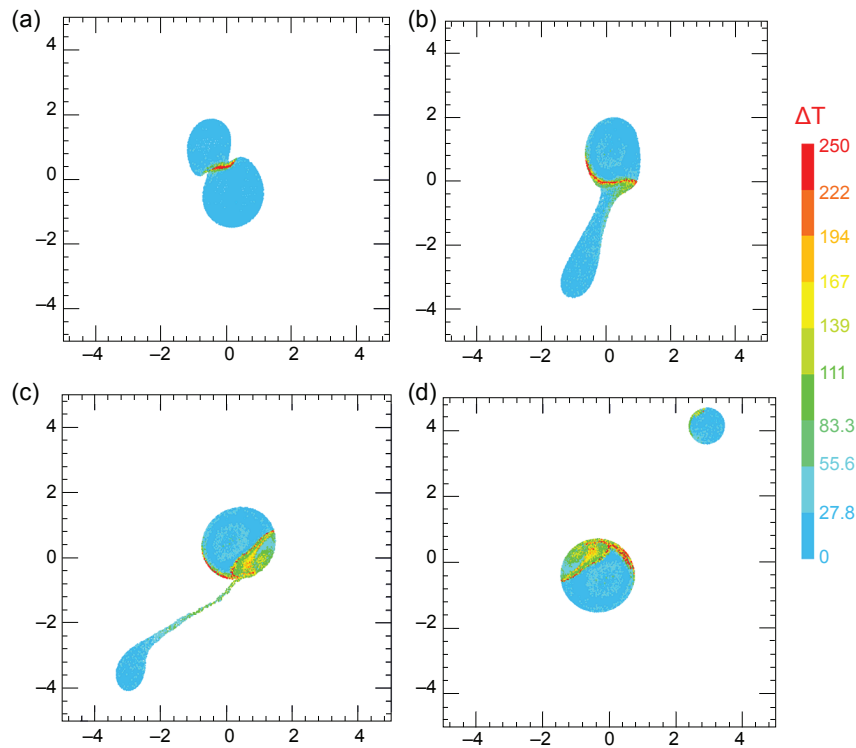

**Fig. 7.** SPH simulation of a potential Pluto-Charon-forming impact that produces an intact Charon. Two non-rotating, undifferentiated/uniform composition bodies undergo an oblique, $v_i = v_{esc}$ collision, with the impactor containing 30% of the Pluto system mass **(a)**. A substantial portion of the impactor remains gravitationally bound after the initial collision **(b)**, and it achieves a bound, stable orbit about the final planet due to gravitational interactions with the distorted shape of the planet **(b),(c)**. The final moon has q = 0.12, an orbital eccentricity of e = 0.5, and a periapse exterior to the Roche limit **(d)**. Color indicates the increase in temperature due to the impact per color bar. From *Canup* (2005).

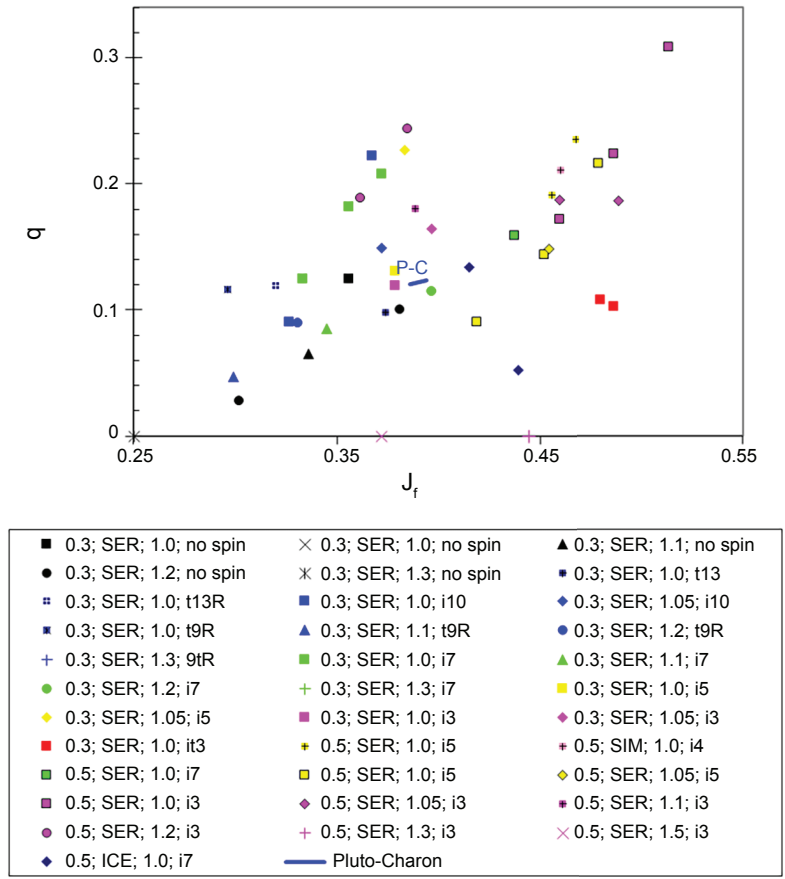

**Fig. 8.** Satellite-to-planet mass ratio (q) from SPH simulations of impacts between uniform-composition bodies that produced intact moons. Labels in the legend provide the same data as in Fig. 6; ICE = uniform ice composition, and "R" after the spin period corresponds to a retrograde pre-impact spin (i.e., a pre-impact spin vector that is anti-aligned with the collision angular momentum vector). Figure from *Canup* (2005), modified to show post-New Horizons values for the Pluto-Charon system (bright blue solid line).

Impacts between the partially differentiated bodies considered in *Canup* (2011) were able to produce both an intact Charon and a much lower-mass disk whose composition ranged from similar to that of the colliding bodies (i.e., 10% pure ice by mass) to 100% ice (Fig. 10). An immediate question was then how the expected radial extent of this dispersed disk compared with the current small moon orbital radii.

A low-velocity, Charon-forming impact produces disk debris with minimal vapor on initially high-eccentricity orbits (e.g., *Canup*, 2005, *Sekine et al.*, 2017). Physically, each SPH disk particle represents a distribution of material with much smaller sizes than can be resolved with SPH. After the impact, orbiting disk debris would first undergo mutual collisions at velocities high enough for rebounding and/or fragmentation. Such collisions would dissipate energy and circularize debris orbits. Once collisions caused the relative velocities of debris to be damped to sufficiently low values, accumulation of moons from the debris could begin. Given this expected evolution, it is common to estimate the orbits to which disk debris would collisionally relax prior to moon accretion by computing an equivalent circular orbit that has the same angular momentum as each initial SPH disk particle, i.e, with $a_{eq} = a(1-e^2)$, where a and e are the SPH particle's post-impact semimajor axis and eccentricity and $a_{eq}$ is the expected semimajor axis when moon accretion begins. The radial extent of a low-mass disk can be then estimated by computing the maximum value, $a_{eq,max}$, among all the disk particles found in an SPH simulation. This value is best interpreted as the distance at or beyond which a mass comparable to that of a single SPH particle is found (e.g., *Canup and Salmon*, 2018).

For simulations with $10^5$–$10^6$ particles (whose individual particle masses are $\sim 10^{19}$–$10^{20}$ g), the estimated outer "edge" of the low-mass disk produced in the *Canup* (2011) simulations was between 5 and 30 Pluto radii ($R_P$) (Fig. 10c), substantially interior to outermost Hydra, which orbits at

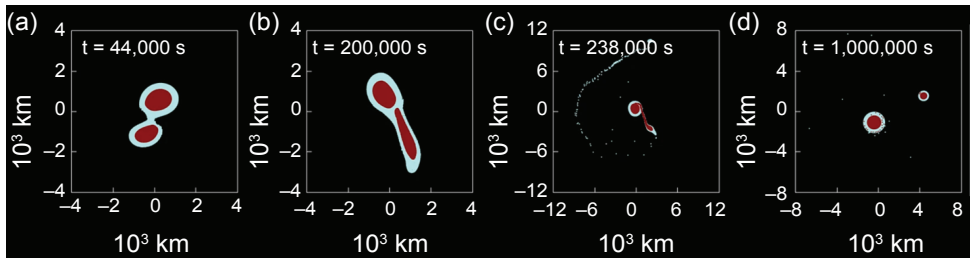

**Fig. 9.** The highly oblique collision of two differentiated bodies with 50% ice (blue) and 50% basalt (red) by mass yields an intact Charon analog with q = 0.13. From *Arakawa et al.* (2019).

~55 $R_P$. It is important to note that these SPH particle masses are comparable to masses of the small satellites themselves, rather than true debris out of which the moons might accrete. Further investigation is warranted to determine if a higher-angular-momentum tail of debris might exist in higher-resolution simulations.

Thus, a giant impact between partially differentiated progenitors appears capable of producing an intact massive Charon, together with a much less massive dispersed disk. The composition of the low-mass, dispersed disk varies across different cases: It can contain a similar bulk composition to that of the impactor or can be completely made of ice. The latter case offers a compelling explanation for the inferred ice-rich compositions of Pluto's small moons, which differ substantially from the darker and more rock-rich compositions inferred for small KBOs. However, the outer edge of the low-mass disk in simulations to date is far too small to directly yield the extended outer moon orbits. On face, this mismatch requires some type of outward migration to produce the small moons from the same impact that produced Charon, a topic to which we now turn.

## 4. ORIGIN OF THE SMALL MOONS

Improved measurements and limits on the masses and orbits of Pluto's outer four small moons have provided new clues for origin models. In this section, we first summarize the uncontentious satellite characteristics (see also the chapter by Porter et al. in this volume), and subsequently review varied theoretical models for their formation. Given the current evidence in favor of a giant impact origin for Charon, we only discuss models that appear consistent with this picture. Current small moon formation models fall into four categories: (1) The moons form in a compact circumbinary debris disk produced by a Charon-forming impact, and then are migrated outward to their present location as Charon's orbit tidally expands; (2) the moons form *in situ* in a large circumbinary debris disk produced by the Charon-forming impact; (3) the moons accrete from the collisional debris of heliocentric material captured long after a Charon-forming impact; and (4) hybrid models in which the moons formed *in situ* (or nearly so) from debris generated by collisions between Charon-impact debris and one or more heliocentric bodies. Despite the abundance of new data provided by New Horizons and many creative theoretical concepts, a coherent model for the origin of the Pluto-Charon circumbinary satellite system remains elusive.

### 4.1. Properties of the Small Moons

The four moons comprise a tiny mass fraction (on the order of $10^{-6}$) of the Pluto-Charon binary. The two larger satellites, Nix and Hydra, have effective radii of approximately 40 km, while smaller Styx and Kerberos are only roughly 10 km in radius. The satellite orbital properties have been traced for over a decade using Hubble Space Telescope (HST) data. All moons have low-eccentricity, low-inclination orbits with respect to the binary and they reside, near, but not in, N:1 mean-motion resonances with the binary (with N = 3, 4, 5, and 6), and thus also near first-order resonances with each other. *Showalter and Hamilton* (2015) identified possible three-body resonances among the satellites, based in part on observations that placed the resonant angle near the expected value for resonant libration. Mean-motion resonances are unlikely given current estimated satellite masses and the low binary eccentricity (*Brozović et al.*, 2015; *Jacobson et al.*, 2019); both drive the resonant widths — the range of period ratios in which libration occurs — toward zero. Future planned observations will confirm whether the resonant angle is circulating (non-resonant) or librating (e.g., *Showalter et al.*, 2019). Accounting for the satellites' near-resonant orbits presents a great challenge to theoretical models. Resonant chains can be the hallmark of dissipative formation processes and differential orbital migration (e.g., *Malhotra*, 1991; *Peale and Lee*, 2002), but such tight dynamical packing, especially around a binary, is difficult to achieve because it is vulnerable to orbital instabilities (e.g., *Sutherland and Kratter*, 2019).

The small moons all have high estimated albedos from ~0.6 to 0.8, brighter than the mean geometric albedo of Charon, which is ~0.4 (*Weaver et al.*, 2016). High small moon albedos were first predicted from the dynamical models of *Youdin et al.* (2012), who found that lower albedos, which required larger moon masses for consistency with HST data, would lead to dynamical instability. Subsequent orbital analyses have confirmed that stability requires high albedos and low moon masses, and in the case of Hydra, have pushed its estimated mass to even lower values (*Brozović et al.*, 2015; *Kenyon and Bromley*, 2019; *Jacobson et al.*, 2019).

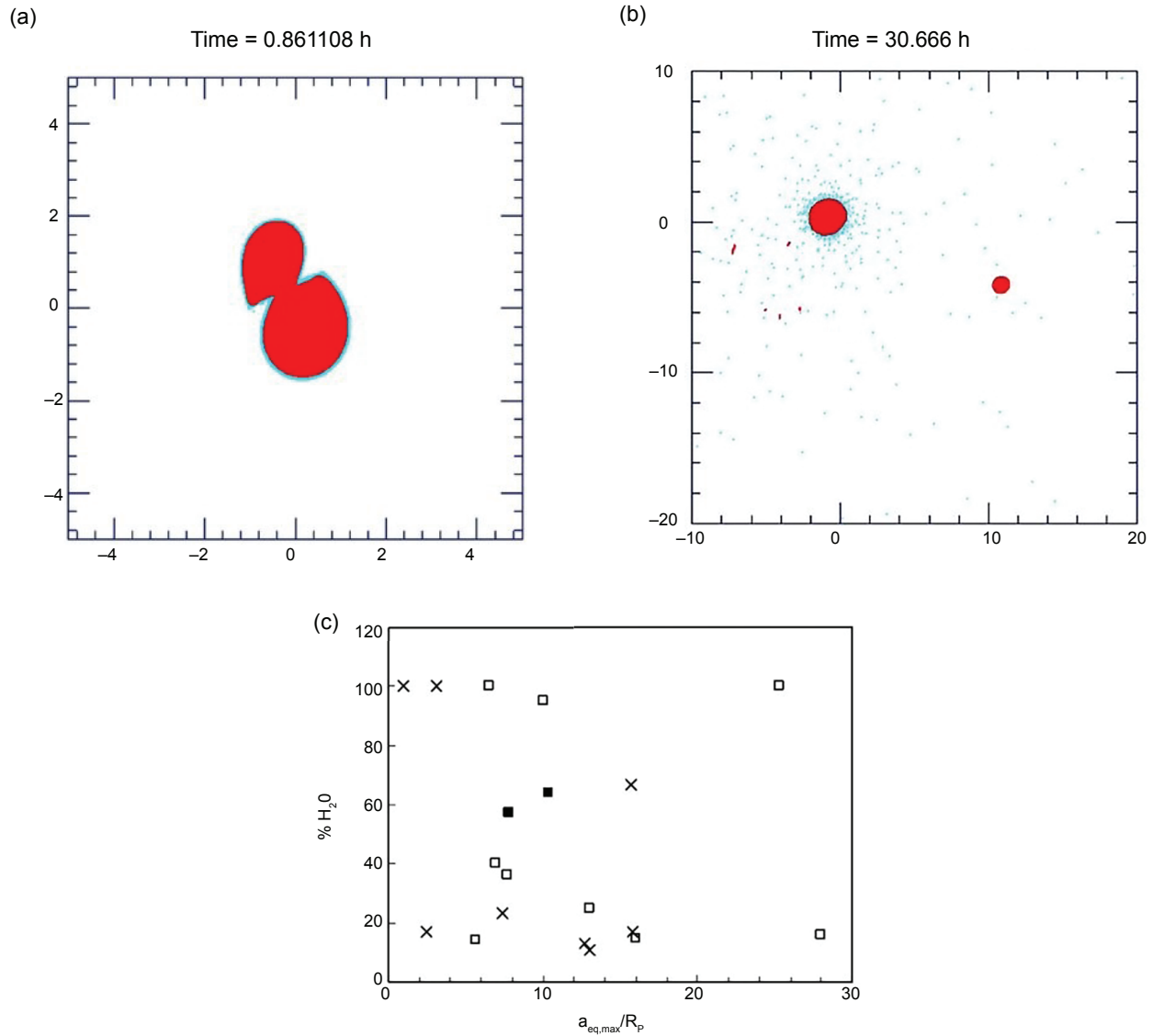

**Fig. 10.** **(a),(b)** Collision between two partially differentiated bodies (90% hydrated silicate, 10% ice shells) leads to the formation of an intact Charon-analog and an extended low-mass disk. **(c)** Composition and outer edge of the low-mass disk. Squares correspond to cases with $\gamma = 0.3$, while ×'s indicate cases with $\gamma = 0.5$. For comparison, the progenitor bodies (both target and impactor) had 10% ice by mass. From *Canup* (2011).

The small moons' high albedos imply bright, ice-dominated surfaces, consistent with spectral measurements by New Horizons (*Cook et al.*, 2018). As discussed in section 2.2, it seems probable that the small moons' bulk compositions are also predominantly icy. However, this currently remains an inference, because the small moon masses and bulk densities are unconstrained by New Horizons data alone (*Weaver et al.*, 2016). Ongoing work to combine small moon shape models (*Porter et al.*, 2019) with improved limits on their masses derived from stability analyses and both HST and New Horizons data (*Jacobson et al.*, 2019) should provide bulk density estimates in the near future. For ice-rich bulk compositions and likely porous interiors, one expects the small moon densities to be $\leq 1$ g cm$^{-3}$ (e.g., *McKinnon et al.*, 2017; *Jacobson et al.*, 2019).

**4.2. Giant Impact Formation and Resonant Expansion of Small Moons**

Given that Charon's orbit is thought to have tidally evolved outward from a few Pluto radii to its current distance at ~17 $R_P$, and that the small moons orbit near N:1 resonances with Charon, it was initially proposed that the moons accreted from Charon-forming impact debris, and were then captured into mean-motion resonances and driven outward to their current locations as Charon's orbit tidally expanded (*Stern et al.*, 2006). The difficulty with this otherwise compelling idea is that, due to the large extent of Charon's radial migration, capture into mean-motion resonances would drive the small moon eccentricities to values so high as to cause destabilization of the small moon orbits. In analogy to excitation of Pluto's current ~0.25 eccentricity from just a ~15% expansion of Neptune's orbit, the much greater post-giant impact tidal expansion of Charon's orbit by a factor of 4 or more would drive objects captured in mean-motion resonances to e > 0.5, destabilizing the resonances and causing mutual collisions, ejection, and/or accretion of the small moons by Charon (*Ward and Canup*, 2006; *Cheng et al.*, 2014b). If such an instability occurred, however, it is possible that the debris could seed a second generation of satellites (see section 4.5).

*Ward and Canup* (2006) proposed a resonant migration model in which the small moons were instead captured into co-rotation resonances with Charon. These resonances exist if Charon's orbit is eccentric, and are located at similar locations as the mean-motion resonances. Co-rotation resonances, which are associated with confinement of Neptune's ring arcs (*Goldreich et al.*, 1986), are notable because they do not increase a trapped object's eccentricity as the perturbing body's orbits expands. In the *Ward and Canup* (2006) model, the current small moons represent a small fraction of an impact-produced debris disk that became trapped into co-rotation resonances with an eccentrically orbiting Charon. So long as Charon maintained an orbital eccentricity as it tidally evolved, the small moons could be driven outward to their current locations on near-circular orbits, with the co-rotation resonances ultimately disappearing as Charon approached the dual-synchronous state and its orbit circularized. The requirement of an eccentrically orbiting Charon is supported by predictions of impact simulations that yield an intact Charon (*Canup*, 2005, 2011; *Arakawa et al.*, 2019) (section 3.5), although maintaining an eccentricity during its subsequent evolution requires a specific range of tidal parameters (e.g., *Cheng et al.*, 2014b). A successful prediction of this model is that the material thrown out to the largest orbital radii could be dominated by ices (*Canup*, 2011) (Fig. 10c), which is consistent with the general increase of albedos from Charon to Hydra (e.g., *Stern et al.*, 2018).

Despite their appeal in terms of avoiding eccentricity growth, even the co-rotation resonances seem to ultimately lead to instability of one or more satellites. *Lithwick and Wu* (2008) argued that the critical eccentricities required for Charon to trap Nix and Hydra are mutually exclusive, with Nix's capture in the 4:1 co-rotation resonance requiring that Charon's eccentricity was $e_C < 0.024$, while Hydra's capture in the 6:1 only occurs for $e_C > 0.04$. The inclusion of omitted secular terms in the disturbing function suggests that there might be a small region of allowed eccentricity space ($e_C \sim 0.05$) at which both captures can occur (*Ward and Canup*, 2010).

In addition, the above analytic models rely on approximations of perturbation theory that may not capture the full system behavior, especially for the large Charon-to-Pluto mass ratio [e.g., see *Mardling* (2013) for expansions valid at arbitrary mass ratios]. Indeed, numerical integrations that include more physics only exacerbate the resonant expansion model's difficulties. *Cheng et al.* (2014b) find that the inclusion of the effects of Pluto's oblateness (i.e., its $J_2$) destabilizes all resonant capture models that they explored for a range of Charon eccentricities. Including the masses of Nix and Hydra in numerical integrations also destabilized outward migration scenarios. In the *Cheng et al.* (2014b) integrations, Nix is always the source of the instability. This is not surprising given that when Charon's eccentricity exceeds ~0.15, Nix's orbit is dynamically unstable on timescales of hundreds of orbits due to overlap of the close-in N:1 resonances (e.g., *Holman and Wiegert*, 1999; *Smullen and Kratter*, 2017). Notably these models focus primarily on Nix and Hydra, and do not claim to place Styx or Kerberos in their current locations either. We discuss these dynamical constraints in more detail in section 4.6.

### 4.3. *In Situ* Formation from the Charon-Impact Debris-Disk

A key motivation for the outward resonant transport model was that giant impacts tend to produce debris disks that are more compact than the current small moon orbits (e.g., *Canup*, 2011) (Fig. 10c), so some type of spreading mechanism seems required. *Kenyon and Bromley* (2014) suggest that impact-generated debris could spread outward to a distance comparable to Hydra's current orbit prior to small moon accumulation, depending on the balance between secular perturbations by an initially eccentrically orbiting Pluto-Charon binary and collisional damping among debris particles. The small moons would then accumulate near their current locations. One issue with such a model is that the collisional spreading mechanism would tend to deactivate as moons accreted and the collision rate decreased, and whether debris could expand by the needed factor of 2 or more in orbital radius before substantial accumulation took place is unclear. Furthermore, the *Kenyon and Bromley* (2014) model predicts a population of small moons and particles orbiting beyond Hydra that were not seen by New Horizons.

Alternatively, it is possible that a Pluto-Charon forming impact could directly produce a low-mass debris disk that is much more extended than seen in prior SPH simulations. Numerical resolution in the *Canup* (2011) simulations appeared adequate to resolve the early estimated total mass of the small moons (*Tholen et al.*, 2008). However, moon mass estimates subsequently decreased by about an order of magnitude (e.g., *Brozović et al.*, 2015), suggesting that their mass was not well resolved by the 2011 models. Studies of low-mass debris disks produced by potential Phobos-Deimos-forming impacts with Mars (*Canup and Salmon*, 2018) show that when simulation resolution is increased, the estimated outer debris disk "edge" typically increases as well, as a lower-mass, higher specific angular momentum component is increasingly resolved.

A challenge for *in situ* formation models is explaining the near-resonant orbits of the small moons. *Smullen and Kratter* (2017) and *Woo and Lee* (2018) explored the behavior of test particles near the current small moon locations in response to Charon's tidal migration. Both studies found that test particles are either evacuated from resonance regions or have too high eccentricity to be consistent with the present-day satellite orbits near N:1 resonances. Note that both studies neglect collisional damping; *Smullen and Kratter* (2017) included no damping, while *Woo and Lee* (2018) damp the test particles to cold orbits prior to Charon's tidal expansion, but not during. The latter scenario is akin to assuming that satellite formation is complete prior to Charon's expansion and leaves behind insufficient debris to damp the satellites. Models with concurrent accretion and collisional damping (but with a different source of debris and without Charon's orbital evolution) do not appear to form the moons near resonances either, as discussed in section 4.4.

A recent novel version of this scenario imagines that the satellites form *in situ* from impact debris, but that the debris is generated by a separate impact on Charon (*Bromley and Kenyon*, 2020). By requiring that the impact occur after the binary has evolved to its present-day orbit, this model avoids the complications of resonant expansion. While the required collision is unlikely, it is not prohibitively so, at least for oblique impacts, which would allow a 30–50-km body to generate sufficient debris. Such a model still does not directly explain the satellite's emplacement near, but not in, resonance with the Pluto-Charon binary; this is a concern given low resonant widths once the binary has circularized.

### 4.4. Moon Origin from Capture of Heliocentric Planetesimals

*Pires Dos Santos et al.* (2012) explored a model in which the satellites form from planetesimals that are captured, disrupted, and reaccreted into the current generation of satellites. They calculated the capture probability for a variety of heliocentric disk models and found that while temporary capture of protosatellites is possible, they tend to be rare and short-lived. To increase the lifetime of the material in orbit, and to ultimately generate low-eccentricity, low-inclination bodies, they considered the probability that captured objects are collisionally disrupted. They concluded that the collisional timescales are longer than the captured satellite lifetimes, thus inhibiting long-term survival of sufficient mass to form the current satellite system. The Pires Dos Santos et al. model did not address the near-resonance moon configurations, and does not offer a clear rationale for their high albedos, since captured material would be expected to have a composition similar to that of small (and much darker) KBOs.

### 4.5. Hybrid Models: Charon Impact Debris Interacts with Heliocentric Debris

The apparent failure of both the resonant expansion and heliocentric capture models has led to the development of hybrid approaches. *Kenyon and Bromley* (2014) and *Walsh and Levison* (2015) considered the outcome of interactions between debris with heliocentric origins and leftover debris or satellites from a Charon-forming impact. Leftover material could radially spread, collisionally damp, and then coalesce into the satellites *in situ*, or at larger separations followed by inward migration. An appealing aspect of these models is that they rely on inward, dissipative movement of either the debris or the satellites themselves, which may allow for the production of resonant chains and/or resonant capture. As with other models, difficulties arise when detailed treatments are considered.

*Kenyon and Bromley* (2014) proposed that debris from the giant impact spreads through interactions with other captured objects in the Pluto-Charon system. In particular, they argued that captured bodies that generate debris pre-date the Charon-forming impact. The old and new debris collide, spread beyond the orbit of Hydra, and then subsequently coalesce into an array of satellites. In their model, these objects can migrate inward and finish near the present-day moon orbits, although they don't show a special preference for the known near-N:1 resonances.

Although aspects of the model are compelling, the omission of some important physics might limit its applicability. Most crucially, many of the spreading and collisional calculations are done in the absence of perturbations from the Pluto-Charon binary. Stirring from the binary can change satellite accretion timescales (*Bromley and Kenyon*, 2015; *Silsbee and Rafikov*, 2015). Although the simulated growth and subsequent evolution of the satellites are assumed to occur after the binary's tidal expansion, the evolution of the initial debris during the expansion is not addressed. In the absence of strong collisional damping, the majority of the debris from the Charon-forming impact could be removed by tidal evolution. If Charon's eccentricity reaches the values expected in current tidal models (*Cheng et al.*, 2014b), some or all of the satellites except Hydra become dynamically unstable on their present-day orbits (*Smullen and Kratter*, 2017). As mentioned above, this model also predicts the growth and survival of small satellites exterior to Hydra (e.g., *Kenyon and Bromley*, 2019); such satellites have not been seen to date (section 2.3).

*Walsh and Levison* (2015) considered the evolution of debris from one or more disrupted outer satellites that formed as a result of a Charon-forming impact. Such disruptions could occur due to interactions with other bodies in the debris disk, through orbital excitation during Charon's tidal evolution or by interactions with heliocentric impactors. The latter is the primary case tested by Walsh and Levison. For consistency with the ancient surfaces of the current moons, any disruptions would need to occur within the first few hundred million years of solar system history. Such early disruption(s) appear likely, due to both the higher primordial flux of heliocentric objects and rapid instability timescales in the Pluto-Charon system. The presence of pre-existing material in the Pluto-Charon disk potentially averts the collision timescale problem identified by *Pires Dos Santos et al.* (2012), although the probability of the aforementioned disruption event is not calculated, and it would depend on the initial conditions in the primordial solar system planetesimal disk.

*Walsh and Levison* (2015) simulate the collisional evolution (including fragmentation and accretion) of an initially eccentric ring of debris generated by a collision between a bound satellite and a heliocentric impactor across a range of different static Pluto-Charon orbits. Similar to *Kenyon and Bromley* (2014), a promising aspect of this model is the ability to launch debris into higher-angular-momentum, stable orbits through the combination of perturbations by Charon and collisional damping. Although the debris rings can successfully damp and reaccrete into satellites, they do not tend to settle into resonant or near-resonant N:1 orbits. The omission of Pluto-Charon orbit evolution in these calculations is a limitation that leaves open possible avenues for more successful satellite emplacement. Although the tidal evolution timescale is slow compared to the accretion timescale, it is possible that subsequent shifts in Charon's semimajor axis and eccentricity might allow debris to be ratcheted into larger orbits through multiple disruptions, if the disruptions occur on a timescale short compared to the ~10-m.y. time for Charon's tidal evolution.

The expected composition of satellites formed in hybrid models is not clear. Pluto's small moons are strikingly brighter than typical KBOs of the same size (*Weaver et al.*, 2016). While predominantly ice-rich dispersed debris can result from a Charon-forming impact (*Canup*, 2011), it would be difficult to preserve such compositional gradients if substantial heliocentric material were added to the small moons.

### 4.6. Circumbinary Dynamics and the Mystery of the Near-Resonant Configuration

A primary unsolved problem in the origin of the small satellites is their packed orbital configuration. The lack of identifiable librating resonance angles in the system today is not, however, too surprising. The N:1 resonances with the binary are all exceedingly narrow (although nonzero) given Charon's current circular orbit (*Mardling*, 2013; *Sutherland and Kratter*, 2019). Moreover, the small masses and low eccentricities of the satellites also produce very narrow libration zones for any first-order satellite-satellite resonances. Since the absence of libration today can be attributed to Charon's circular orbit, it is natural to cite the orbital configuration as evidence that the system acquired these period ratios when Charon was still eccentric, and thus resonant libration more likely. This constraint pins the formation of the satellites to the first few to 10 m.y. after a Charon-forming impact (e.g., *Cheng et al.*, 2014). This timing is consistent with the aforementioned small moon origin models, as they either rely on Charon's outward migration or on a high flux of heliocentric impactors most consistent with the early solar system; an early formation is also consistent with the ancient cratered surfaces of the small moons (e.g., *Robbins et al.*, 2017; *Singer et al.*, 2019).

However, recent work by *Sutherland and Kratter* (2019) shows that multi-object resonant configurations are tenuous even if the binary's mutual orbit is eccentric. Chaos leading to orbital instability can occur due to N:1 resonance overlap, N:1 and first-order satellite-satellite resonance overlap, and satellite-satellite resonance overlap. Stability is particularly tenuous due to the combination of the high forced eccentricities due to the binary and resonance splitting (the breakup of one resonance into many due to different precession rates driven by non-Keplerian potentials). Unlike in satellite systems surrounding single bodies, in which resonance splitting driven by the primary's $J_2$ can increase stability, the substantial secular forced eccentricity generated by the binary broadens libration zones so that many different combinations can overlap. For example, accounting for the forced eccentricity from Pluto-Charon, the Kerberos-Hydra pair would be unstable for $e_C > 0.18$ because the 5:6 and 6:7 resonances overlap. At $e_C \sim 0.14$, the 5:6 overlaps with the 11:13. Eccentricity pumping by the binary N:1 resonances exacerbate the issue, especially for the inner satellites. Strong collisional damping could forestall some of the aforementioned instabilities, but it is unclear whether enough mass exists in the system to both form the satellites and damp them as an eccentrically orbiting Charon tidally evolves.

### 4.7. Future Outlook

While the wealth of data from New Horizons has not yet yielded an obvious solution to the "small moon problem," we have identified several promising directions for future investigations. First, higher-resolution simulations of the post-impact disk are crucial for constraining the initial conditions for moon formation. Such simulations must provide mass resolution that extends below the current estimated satellite masses. Second, more thorough investigation of scenarios in which the observed satellites are not first-generation satellites, but rather have been reprocessed through collisions either with or without the incorporation of pre-existing heliocentric material, should be explored. Finally, future dynamical models of satellite accretion and migration should simultaneously incorporate the tidal migration of Charon and collisional damping within the debris disk.

## 5. IMPLICATIONS AND OPEN ISSUES

In this final section, we discuss implications of giant impact models for the environment at the time of Pluto system formation, the possible preservation of a collisional "family" of ejecta from a Pluto-system-forming impact, the origin of Pluto and Charon volatiles, and implications for other KBO binary systems. We finish by highlighting key topics for future work.

### 5.1. Implications of an Impact Origin for the Transneptunian Environment at the Time of Pluto System Formation

A general result of all impact models that can produce massive Charon is the requirement of a very low collision velocity that is within 20% of the mutual escape velocity of the progenitor bodies (*Canup*, 2005, 2011; *Arakawa et al.*, 2019). For similarly sized progenitors as needed to account for the system's high angular momentum (with $\gamma = 0.3–0.5$), $v_{esc} \approx 1$ km s$^{-1}$. The impact velocity is $v_i^2 = v_{esc}^2 + v_{rel}^2$, so that requiring $v_i \leq 1.2\, v_{esc}$ implies a relative velocity at large separation of $v_{rel} \leq 0.7$ km s$^{-1}$. This is substantially smaller than the current Kuiper belt velocity dispersion, which is ~1–2 km s$^{-1}$. It is thought that in the primordial epoch — in particular before the migration of Neptune dynamically excited the transneptunian region — relative velocities among large ~10$^3$-km bodies would have been substantially lower than they are in the current Kuiper belt. This seems consistent with conditions needed to have accreted such large bodies from a population of ~10$^2$-km planetesimals produced by the streaming instability and background remnant pebbles (e.g., *Morbidelli and Nesvorný*, 2019). Thus, the low collision velocity needed for a Pluto-Charon forming impact seems most consistent with the pre-outer planet migration era. As noted in section 2.5, a later origin of the binary after resonance capture cannot be ruled out based on Pluto's heliocentric orbit alone, because the system could have been recaptured into resonance after a giant impact (e.g., *Dobrovolskis et al.*, 1997). However, this requires "fine tuning" of conditions to both achieve a low relative velocity between the progenitors once Neptune's migration and dynamical excitation of the transneptunian region have begun, and to allow recapture of the post-impact system into the 3:2 resonance.

It is illustrative to also consider the general likelihood of a collision between two $R_e \sim 10^3$-km-class bodies. A simple "particle-in-a-box" estimate for the time between collisions among any of the objects within a swarm of N such bodies orbiting across a region extending in orbital radius from a to (a + Δa) gives

$$t_{col} \sim \frac{2\pi\, a\, \Delta a\, H}{N^2 \pi\, (R_e + R_e)^2 f_g v_{rel}} \quad (7)$$

where $H \sim v_{rel}/\Omega$ is the swarm's vertical thickness, $\Omega(a)$ is orbital heliocentric frequency, and $f_g = [1 + (v_{esc}/v_{rel})^2] \geq 2$ is a gravitational focusing factor. For a pre-Neptune migration population with a = 20 AU, Δa = 10 AU, and $R_e = 10^3$ km, one collision occurs every

$$t_{col} \sim \text{few} \times 10^6 \left(\frac{10^3}{N}\right)^2 \left(\frac{10}{f_g}\right) \text{y} \quad (8)$$

The likelihood of a primordial Pluto-Charon forming collision is vanishingly small if there had only been an initial population of N ~ 10 Pluto-scale bodies, as predicted by some hierarchical accretion models (e.g., *Kenyon and Bromley*, 2012; *Schlichting et al.*, 2013), unless $f_g$ was extremely large. However, if there were initially N ~ few × 10$^3$ Pluto-scale bodies, $t_{col}$ is on the order of 10$^6$ yr or less, depending on $f_g$. Appealingly, this is consistent both with the number of Pluto-class bodies needed to account for Neptune's "grainy" migration and the observed number of resonant vs. non-resonant KBOs [N ~ 2000–4000 (*Nesvorný and Vokrouhlický*, 2016)] and with large impacts in this region occurring prior to an outer-planet migration within the first tens of millions of years after nebular dispersal. The latter timing avoids a destabilization of the inner terrestrial planets associated with a later migration (e.g., *Agnor and Lin*, 2012; *Kaib and Chambers*, 2016) (see section 2.5).

However, the efficiency of implantation of original disk bodies into the Kuiper belt in dynamical models is low, ~10$^{-3}$ (*Nesvorný and Vokrouhlický*, 2016) (see section 2.1). With N = 4000, equation (8) implies about 300 giant impacts within ~50 m.y. for $f_g$ = 10. There would then be a substantial probability that a single Pluto-sized body that had experienced a giant impact would also be retained. It is possible that in the cold classical belt, relative velocities could have been very low, leading to $f_g \approx 10^2$, making all Pluto-scale bodies likely to have experienced a giant impact within 50 m.y. Having $f_g \sim 10–10^2$ requires $v_{rel} \sim (1/10)$ to $(1/3)\, v_{esc}$, which in turn implies a Pluto-Charon forming impact with $v_i < 1.1\, v_{esc}$, consistent with (but independent from) requirements from impact simulations.

Thus, a low-velocity collision appears needed both to make a Charon-sized satellite and for such a giant impact to have been probable in the first few tens of millions of years of the solar system's evolution. This timing (Fig. 11) meshes well with that implied by current models for outer planet migration and Kuiper belt formation (e.g., reviews by *Nesvorný*, 2018; *Morbidelli and Nesvorný*, 2019).

### 5.2. Possibility of a Pluto Collisional Family

Additional evidence supporting a giant-impact scenario for the formation of Pluto-Charon (and potentially of the small moons as well) could be the identification of collisional family members. Even a collisional family formed prior to the solar system rearrangement that caused Pluto's heliocentric migration could have survived this migration (*Smullen and Kratter*, 2017).

To date, a single transneptunian collisional family has been identified, for Haumea (*Brown et al.*, 2007). Family members were singled out based on orbital (a low dispersion velocity, <140 m s$^{-1}$, among family members) and surface properties (icy) similar to Haumea's, even though Haumea's orbital eccentricity is higher than those of identified family members, presumably having been increased by gravitational interactions with Neptune since the time of impact.

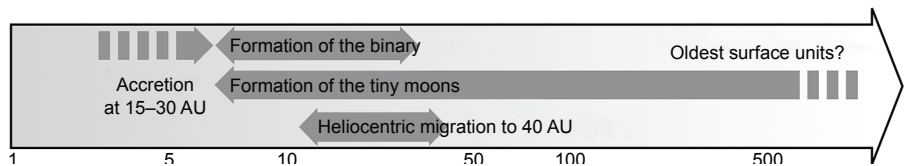

**Fig. 11.** Tentative timeline of origin events in the Pluto system. Accretion and heliocentric migration are discussed in section 2, the formation of the binary in section 3, and that of the tiny moons in section 4.

Identifying a collisional family for Pluto could be more challenging. Typically, the dispersion velocity of a collisional family would be expected to be comparable to Pluto's escape velocity [$v_{esc} \approx 1.2$ km s$^{-1}$ (e.g., *Schlichting and Sari*, 2009)], implying that family members could span much of the dynamical space of observed transneptunian objects (Fig. 12). In their simulations of graze-and-merge collisions as a possible origin for the Haumea system, *Leinhardt et al.* (2010) showed that for this type of impact (which is similar to impacts shown in Figs. 5–6 that form massive disks), most escaping ejecta have velocity dispersions <0.5× the primary's escape velocity, and a substantial portion have dispersion velocities ≤0.2 $v_{esc}$. *Smullen and Kratter* (2017) considered material ejected post-impact due to dynamical interactions between the Pluto-Charon binary and a low-mass disk of material [similar to conditions found in *Canup* (2011)] as Charon's orbit tidally expands. The low-mass disk is described with test particles, so mutual collisions are neglected. Of the disk material that is ejected from the binary, they find that on the order of 10% remains on stable heliocentric orbits after 1.5 G.y., even when effects of a smooth migration of Neptune are included. The majority of survivors are trapped in the 3:2 resonance as a population of Plutinos, with characteristic velocity dispersions ~100 to 200 m s$^{-1}$ (*Smullen and Kratter*, 2017); these velocities would likely be modestly reduced if collisions among the ejecta had been included. The omission of the effects of grainy migration likely enhances resonance capture rates (*Nesvorný and Vokrouhlicky*, 2016). A Pluto family would thus have a low-velocity dispersion, but its orbital properties could be difficult to distinguish from those of other resonant KBOs.

A more distinguishing feature of a Pluto family could be its composition. The composition of family members may preferentially reflect that of the progenitors' outer layers at the time of impact. This is not as firmly established for Pluto (section 3.3) as it seems for Haumea. Pluto's family could be icy, as suggested by the giant impact simulations of *Canup* (2011) in which the progenitors have an icy veneer, and by the inferred icy compositions of the tiny moons if they formed during this impact. An icy composition would be easiest to identify against the backdrop of generally dark TNOs (*Johnston*, 2018). Alternatively, fragments ejected from progenitors with rock-ice outer layers at the time of the impact (*Desch*, 2015; *Desch and Neveu*, 2017) could have surface properties indistinguishable from those of other TNOs.

### 5.3. Origin of Volatiles and Expected Giant-Impact Heating

The abundances of $N_2$, CO, and $CH_4$ on Pluto and $NH_3$ on Charon are on the whole consistent with a primordial supply as estimated from observations of comets and interstellar sources (*Stern et al.*, 1997; *Glein and Waite*, 2018). To first order, the contrasting presence of CO, $CH_4$, and $N_2$ on Pluto and their absence on Charon is readily explained by Charon's inability to retain these volatiles owing to its lower gravity (*Trafton et al.*, 1988; *Schaller and Brown*, 2007). As such, Pluto and Charon may be two archetypes of the bimodal volatile inventories detected on other large KBOs (*Brown*, 2012). In contrast, ammonia's high miscibility with water (as a polar molecule able to form hydrogen bonds) makes it prone to retention in $H_2O$ ice, as observed throughout Charon's surface, in areas exposed through Pluto's volatile ice veneer, and even on Nix and Hydra (*Cook et al.*, 2018). CO, $CH_4$, and $N_2$ escape could have taken place on Pluto as well, but if this is the case the rates of atmospheric and surface accumulation from a subsurface source exceed those of escape.

Heating of Pluto during a Pluto-Charon-forming impact would have occurred primarily in the region of the initial impact, particularly for the favored low-velocity, oblique collisions that produce Charon intact. Impact simulations predict this region would have been heated by $\Delta T \sim 150$–200 K, leading to local melt pool production (*Canup*, 2005; *Sekine et al.*, 2017). *Sekine et al.* (2017) proposed that if the progenitors had comet-like volatile abundances (notably in $CH_2O$ and $NH_3$), then chemistry in the water melt pool produced by a Charon-forming impact could have led to the formation of Pluto's dark, reddish regions along its equator, and more generally, that such impact-induced chemistry could have caused the color variety observed in large KBOs. For impacts that produce an intact Charon, Charon is only minimally heated by the collision itself, because it largely represents a portion of the progenitor impactor that avoids direct collision with proto-Pluto (*Canup*, 2005). However, if its orbit remains eccentric as it evolves, heating via tidal dissipation becomes important (e.g., *Rhoden et al.*, 2015).

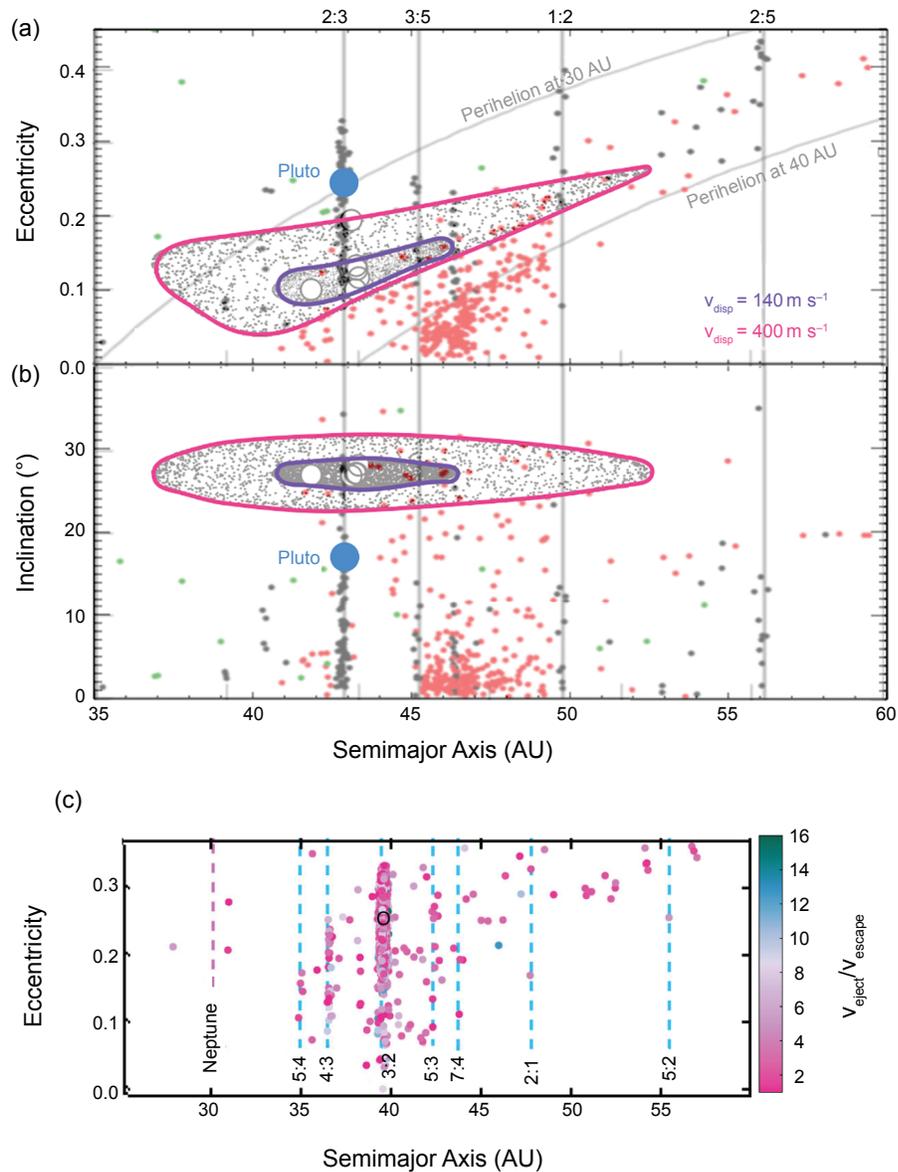

**Fig. 12.** The challenge in detecting Pluto collisional family members based on orbital properties alone. **(a),(b)** The purple and pink outlines represent the space of possible orbital attributes of members of the Haumea collisional family assuming a dispersion velocity of 140 and 400 m s$^{-1}$, respectively, centered on the average position of identified family members. Haumea is the open circle at eccentricity 0.19 and inclination 28°; resonant interactions with Neptune have increased its eccentricity since the time of the collision (*Brown et al.*, 2007). Vertical lines indicate mean-motion resonances with Neptune, with black, red, and green dots indicating resonant, stable non-resonant, and Neptune-encountering KBOs. Modified from Fig. 3 of *Brown et al.* (2007) and Fig. 1 of *Levison et al.* (2008). **(c)** Eccentricity vs. semi-major axis predicted for debris ejected from the Pluto system after 1.5 G.y. of dynamical evolution, including effects of a smooth migration of Neptune (*Smullen and Kratter*, 2017). Color scales with ejection velocity scaled to the escape velocity at Pluto's Hill sphere (note that the latter is different than $v_{esc}$ defined in the main text). Of the ejected debris that survives, most (~60–80%) remains trapped in the 3:2 resonance, with a minority in the 5:3 (few to 5%) and other resonances. Ejecta from the Pluto system that persists in resonances does not display a strong clustering of orbital angles with Pluto. Thus, it could be challenging to distinguish this material from that independently trapped into resonance based on orbital properties alone. Modified from Fig. 7 of *Smullen and Kratter* (2017).

Thus, the Pluto-system progenitors may have accreted volatile-rich, and largely retained volatiles through a hypothesized giant impact. However, except for CO, none of the observed volatiles on Pluto and Charon need be primordial. $N_2$ could result from the oxidation of $NH_3$ and/or organic nitrogen (*Neveu et al.*, 2017). $CH_4$ could be a product of the (kinetically slow) reduction in liquid water of $CO_2$, CO, or organic C (*Shock and McKinnon*, 1993). In the latter case, carbon reactions were partial even on Pluto, whose surface CO would otherwise have been converted to more stable species (*Shock and McKinnon*, 1993; *Glein and Waite*, 2018). Thus, the lower CO/$N_2$ ratio on Pluto relative to most comets (*Stern et al.*, 1997, and references therein) could be explained by post-accretional processes that deplete CO and/or produce $N_2$. Alternatively, it may reflect Pluto's accretion in a $N_2$-rich, CO-poor nebular region that few observed comets sample (*Lisse et al.*, 2020).

### 5.4. Implications for Satellite Origin at Other Dwarf Planets in the Kuiper Belt

The picture advanced in *Nesvorný* (2018), *Morbidelli and Nesvorný* (2019), and section 5.1 envisions thousands of Pluto-class bodies in the 15–30-AU region prior to Neptune's migration, suggesting that giant impacts between such bodies would have been common. Most such objects would have been lost to ejection from the solar system or accretion by the planets. Surviving impact-generated systems would be expected to vary depending on the specifics of individual collisions (e.g., *Arakawa et al.*, 2019). Differences in impact angle and/or velocity between two like-sized dwarf planets can lead to larger ice/rock fractionations (*Barr and Schwamb*, 2016), e.g., leading possibly to Eris-Dysnomia (*Brown et al.*, 2006; *Greenberg and Barnes*, 2008) and Orcus-Vanth (*Brown et al.*, 2010); or outcomes other than a binary (*Brown et al.*, 2006; *Canup*, 2011), such as Haumea, which only has small moons and a collisional family (*Leinhardt et al.*, 2010). Although the orbit, masses, and compositions of Pluto-Charon strongly implicate an impact origin, those of other binary systems could be compatible with alternative origins (Fig. 2). One is capture (*Goldreich et al.*, 2002), e.g., for Eris and much darker Dysnomia (*Greenberg and Barnes*, 2008; *Brown and Butler*, 2018), Orcus-Vanth (*Brown et al.*, 2010), or the eccentric moon of 2007 $OR_{10}$ (*Kiss et al.*, 2019). However, for (at least) dynamically cold, 100-km-class Kuiper belt binaries, the capture mechanisms of *Goldreich et al.* (2002) — dynamical friction from a sea of small planetesimals ("L²s") and three-body encounters ("L³") — appear to be ruled out by the observed distribution of the binaries' mutual orbit inclinations (*Grundy et al.*, 2019a; *Nesvorný et al.*, 2019). A likelier alternative origin is co-accretion (*Nesvorný et al.*, 2010, 2019; *Grundy et al.*, 2019a), e.g., for G!kún‖'hòmdímà-G!ò'é!hú (*Grundy et al.*, 2019b). Possible origins may also depend on formation location (dynamical class). Studies of binary properties generally will help constrain these formation mechanisms.

Simple dynamical arguments can identify the regime of secondary-to-primary mass ratio (q) and orbital separations consistent with impact-produced planet-moon systems that have subsequently tidally evolved (Fig. 13). A probable value for the maximum normalized angular momentum in an impact-produced system is $J \approx 0.45$ (equation (3)), which in combination with equation (2) can be used to calculate the orbital separation for a planet-moon pair that has reached the dual synchronous state as a function of q (solid curve in Fig. 13). However, for small q, tides are too slow for the system to have reached this state over the age of the solar system. A simplified expression (e.g., *Goldreich and Soter*, 1966) for the rate of expansion of a moon's semimajor axis (a) due to tides raised by the moon on the primary, albeit likely conservative for the high orbital eccentricities (e ≥ 0.1) that may result from a binary-forming impact (section 3.5), can be integrated to yield the expected a after t = 4.5 G.y. of evolution

$$\frac{a}{R_p} \approx \left[ 20q \left(\frac{k_2}{Q}\right) \left(\frac{GM_p}{R_p^3}\right)^{\frac{1}{2}} t \right]^{\frac{2}{13}} \quad (9)$$

where $R_p$, $M_p$, $k_2$, and Q are the primary's radius, mass, Love number, and tidal dissipation factor. Assuming a primary density of 2 g cm⁻³, expected separations for $(Q/k_2)$ = 300, 100, and 30 are shown in Fig. 13. Comparison with estimated properties for the large KBOs that have known satellites suggests that most such systems appear compatible with an impact origin (e.g., *Arakawa et al.*, 2019), although the Haumea system appears to require a rather unusually rapid rate of tidal evolution (corresponding to a low primary $Q/k_2$) to have reached its current separation.

Thus, impact-generated systems would be relatively compact, with satellite orbital radii extending out to only ~100 primary radii or less. Such systems would be equally likely to be prograde or retrograde with respect to their heliocentric orbits, but would generally have satellites that orbit in the same sense as the primary's rotation (although see *Rufu et al.*, 2017). Oblique, low-velocity impacts often produce disks (or moons) that are disproportionately comprised of the outer layers of the colliding bodies, so that unusually ice-rich moons are most easily explained by this formation mechanism. However, compositional outcome can depend on impact parameters — e.g., in grazing, low-velocity collisions, much of the impactor core can be incorporated into a moon, leading to similar primary-moon bulk compositions.

### 5.5. Key Open Issues

We conclude by listing some important outstanding issues for future study that could help clarify current uncertainties in Pluto-system origin models. Many relate to the origin of the outer small moons, as this is currently the least well understood element of such models.

- ***Improved giant-impact models.*** In the simplest (and thus most attractive) model, the outer small moons share

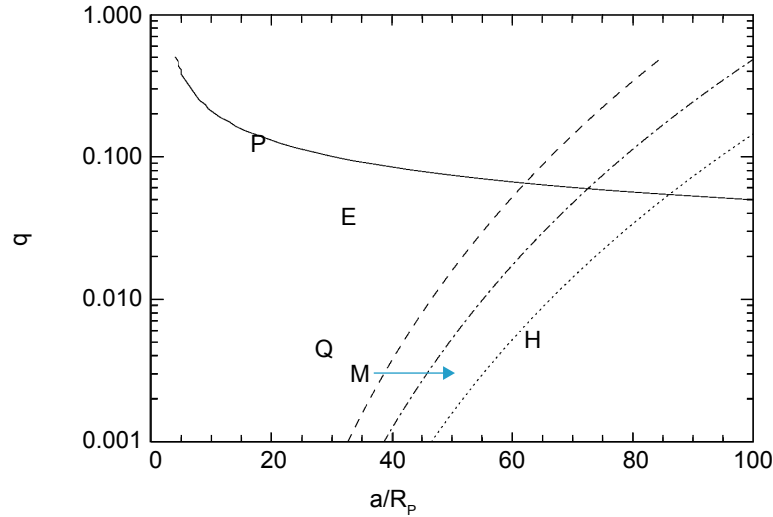

**Fig. 13.** Properties expected for tidally evolved KBO-moon systems that originally formed by a giant impact. The solid curve shows the relationship between the secondary-to-primary mass ratio (q) and the semi-major axis scaled to the primary's radius ($a/R_P$) at which the pair would occupy a dual synchronous state (like that of Pluto-Charon) in the limiting case of the maximum angular momentum that can be imparted by a single impact ($J \approx 0.45$; see equations (2) and (3), and we assume moment of inertia constants for the primary and secondary K = 0.35). Corresponding curves for lower-angular-momentum systems fall beneath this line. A second limit arises from the orbital separation that can be achieved by a pair via tidal evolution over the finite lifetime of the solar system. This is shown as dashed, dot-dashed, and dotted curves for tidal parameters for the primary body of ($Q/k_2$) = 300, 100, and 30, respectively. The region expected for impact-generated systems lies below the solid curve and to the left of the appropriate tidal curve. Approximate properties of KBO-moon systems are indicated by labels (P = Pluto, E = Eris, Q = Quaoar, M = Makemake, and H = Haumea). For Makemake, a blue arrow indicates that only a lower limit on the orbital size is known. Most and perhaps all satellites of large KBOs are potentially consistent with an impact origin.

a common origin with that of the Pluto-Charon binary. To date, a major unresolved challenge is accounting for the large orbital radii of the small moons, which appear much larger than would be directly produced by dispersed debris from a Charon-forming impact. Future work should explore the effects of increased resolution (e.g., *Canup and Salmon*, 2018), varied progenitor interior models (e.g., *Desch*, 2015), and the possible effects of internal strength on impact outcome (e.g., *Davies and Stewart*, 2016; *Emsenhuber et al.*, 2018). The suite of simulations by *Arakawa et al.* (2019) explored impact velocities up to 1.7 $v_{esc}$; given the importance of the currently understood low-velocity constraint for forming Pluto-Charon, investigation of larger speeds to determine if unexpectedly different outcomes occur is warranted.

- **Determination of small moon masses and bulk densities.** Ultimately, estimates of the small moon bulk densities should be obtainable via a combination of New Horizons-derived orbits and shape models, telescopic data, and dynamical stability integrations (e.g., *Porter et al.*, 2019; *Jacobson et al.*, 2019). If these densities remain consistent with the currently inferred ice-rich compositions, this would be a strong indication that the moons originated from material ejected from a Charon-forming collision involving partially differentiated bodies with ice-rich outer layers (e.g., *Canup*, 2011). If their bulk densities instead imply ice-rock compositions more similar to that of Charon, this could be consistent with either an impact origin (Fig. 10c) or with a substantial component of heliocentric material in the small moons. Any substantial moon-to-moon compositional variations would affect origin concepts as well.

- **Detection of remnant signs of an impact origin.** Tidal evolution of Pluto-Charon from an initially more compact state might have led to a fossil oblateness in either body, but this was not observed (e.g., *Moore et al.*, 2016; *Nimmo et al.*, 2017), consistent with the presence of an early Pluto ocean that decoupled the shell from the interior and precluded survival of a fossil bulge (e.g., *Robuchon and Nimmo*, 2011). However, other possible signs of an impact origin may yet exist. It is possible that a collisional family from a Charon-forming impact could survive until the current day. Detection or new upper limits on a collisional family with a composition similar to the unusually bright small moons, together with estimates of dynamical clustering with or without the effect of the giant-planet-induced scattering, could help constrain both the mode and timing of Pluto system formation. Tectonic features on Charon appear most easily explained by a combination of expansion due to ocean freezing and tidal stresses. If these could be used to infer plausible tidal evolution histories (i.e., eccentricity as a function of semimajor axis and/or a as a function of time), or alternatively, histories that would be inconsistent with Charon as observed, this could provide new constraints to origin models (e.g., *Rhoden et al.*, 2015, 2020).

- *More detailed models of small moon formation and orbital evolution.* Unless the small moons expanded in resonance with Charon — which currently does not appear viable across large radial distances — the origin of their near-resonant orbits remains mysterious. However, it has been challenging for models to include all the main relevant processes to small moon assembly, including collisional damping and excitation by a tidally evolving inner Pluto-Charon binary. Future work should attempt to include such additional physics, and consider scenarios in which the current moons undergo only modest migration in resonance. It could also prove helpful to consider whether placing one or two moons in a resonant state (e.g., Nix) might then make the near-resonant states of the other moons probable based on stability arguments alone (e.g., *Walsh and Levison*, 2015).

- *Interior models of progenitors and improved timing of formation constraints.* Impact outcomes are sensitive to the interior state of the colliding bodies. As such, interior models that more fully consider the effects of accretional and radiogenic heating would be valuable. Better limits on the absolute timing of a Pluto-system-forming impact might be obtained by coupling models of the geophysical evolution of progenitors, refined impact models that account for detailed interior structures, and predictions of dynamical models of outer planet migration and excitation. Thermal evolution simulations suggest that the structures of progenitor interiors change significantly within the first tens of million years after accretion (section 3) if accretion takes place 5–10 m.y. after the birth of the solar system (Fig. 4). Determining the sensitivity of impact outcomes to a broader range of progenitor structures would require equations of state for rock-ice mixtures, rather than simply rock, ice, or hydrated silicate equations of state utilized in impact simulations to date.

- *Improved understanding of Kuiper belt size distribution and its implications for small moon origin models.* A lower bound on the age of the Pluto system can be set based on observed crater counts and estimates of the number and size distribution of transneptunian objects through time, which can be used to refine absolute ages of the oldest surface units. On Charon, ~4-G.y.-old surface age estimates from crater counts (*Greenstreet et al.*, 2015; *Stern et al.*, 2015; *Moore et al.*, 2016; *Singer et al.*, 2019) have yet to be reconciled with thermal evolution model predictions of resurfacing due to freezing of a subsurface ocean as recently as 2 Ga (*Desch and Neveu*, 2017). Upcoming deep, automated surveys of the transneptunian region will help firm up our knowledge of the TNO size distribution and how it came to be, enabling us to better place Pluto-Charon in this context. Additionally, the crater record at Pluto implies a surprising lack of impactors with radii ≤1 km (*Singer et al.*, 2019), and the importance of this dearth in background small bodies for small moon origin models that invoke heliocentric contributions should be assessed.

- *Improved census of KBO binaries.* A criticism of the giant impact hypothesis has been the necessity for rather finely tuned collisional conditions to produce the Pluto-Charon system, which lies at the upper end of binary angular momenta that can be produced by a collision. However, if there exists a population of KBO binaries whose properties are consistent with a broader range of impact outcomes, Pluto-Charon could be understood as an easily detectable outlier. As noted in section 5.4, giant impacts may have been common prior to Neptune's migration. At present we can merely state that other large mass ratio systems appear compatible with an impact origin (*Arakawa et al.*, 2019) (Fig. 13). Future work should make improved predictions for the expected mass ratio and semimajor axis distributions of impact-produced binaries that can be benchmarked against bias-corrected KBO samples from the next generation of large surveys.

*Acknowledgments.* We thank W. B. McKinnon and S. A. Stern for helpful discussions and suggestions that helped improve the content of this chapter, and W. B. McKinnon, E. Asphaug, and S. Desch for their helpful and constructive reviews. Support from NASA's New Frontiers Data Analysis program (R.M.C.), the Heising-Simons Foundation and NASA Grant No. 80NSSC18K0726 (K.M.K.), and the CRESST II agreement No. 80GSFC21M0002 between NASA GSFC and Univ. Maryland (M.N.) is gratefully acknowledged.